\documentclass[conference,compsoc]{IEEEtran}

\usepackage[T1]{fontenc}
\usepackage[utf8]{inputenc}
\usepackage{cite}

\usepackage{amsmath,amssymb,amsfonts}
\usepackage{array}
\usepackage{booktabs}
\usepackage{graphicx}
\usepackage{tikz}
\usetikzlibrary{calc,fit}
\usepackage{xcolor}

\usepackage{pgfplots}
\pgfplotsset{compat=1.17}
\usetikzlibrary{pgfplots.groupplots}
\definecolor{ablbase}{HTML}{8FA6C9}   %
\definecolor{ablours}{HTML}{E7AD80}   %
\definecolor{ablgrnA}{HTML}{D6DCB4}   %
\definecolor{ablgrnB}{HTML}{9DC4A8}   %
\definecolor{ablgrnC}{HTML}{9AAD6C}   %
\pgfplotsset{
  ablpanel/.style={
    height=3.0cm,
    axis lines*=left,
    axis line style={black!65, line width=0.4pt},
    tick style={black!65, line width=0.4pt},
    xtick style={draw=none},
    tick label style={font=\scriptsize, /pgf/number format/assume math mode=true},
    scaled y ticks=false,
    yticklabel style={font=\scriptsize, /pgf/number format/fixed, /pgf/number format/assume math mode=true},
    xticklabel style={font=\scriptsize, align=center},
    label style={font=\scriptsize},
    ylabel style={font=\footnotesize, align=center},
    title style={font=\footnotesize, align=center},
    ymin=0,
    clip=false,
  },
}
\tikzset{ablbarlabel/.style={font=\scriptsize, inner sep=1.5pt}}
\usepackage{url}
\usepackage[hidelinks]{hyperref}
\usepackage{pifont}

\usepackage[scaled=0.95]{inconsolata} %
\usepackage{listings}
\lstdefinestyle{eccode}{%
  basicstyle=\ttfamily\footnotesize,
  columns=fullflexible,
  keepspaces=true,
  breaklines=true,
  breakatwhitespace=true,
  breakindent=0pt,
  showstringspaces=false,
  aboveskip=0.7\baselineskip,
  belowskip=0.4\baselineskip,
  morecomment=[n]{(*}{*)},
  commentstyle=\color{black!55}\itshape,
  morekeywords={module,proc,var,return,if,else,while,lemma,equiv,forall,proof,qed,glob,res},
  keywordstyle=\bfseries,
}

\usepackage[most]{tcolorbox}

\definecolor{codebg}{HTML}{F7F8FA}
\definecolor{codeframe}{HTML}{D0D7DE}
\definecolor{codekw}{HTML}{24292F}

\lstdefinestyle{eccodepretty}{
  basicstyle=\ttfamily\footnotesize,
  keywordstyle=\bfseries,
  commentstyle=\itshape,
  columns=fullflexible,
  keepspaces=true,
  breaklines=true,
  showstringspaces=false,
  mathescape=false,
  texcl=false,
  morekeywords={
    module,var,proc,if,then,else,return,true,false,empty,oget,dom, equiv
  },
  literate=
    {<\$}{{<\textdollar}}2
    {<-}{{$\leftarrow$}}2
    {<@}{{$\leftarrow$}}2
    {\/}{{$\vee$}}2
    {`|`}{{$\cup$}}3
}

\newtcblisting{eccodebox}{
  listing only,
  listing engine=listings,
  enhanced,
  breakable,
  colback=codebg,
  colframe=codeframe,
  arc=1.5mm,
  boxrule=0.45pt,
  left=1.5mm,
  right=1.5mm,
  top=1mm,
  bottom=1mm,
  listing options={style=eccodepretty}
}

\newcommand{\system}{ShannonProver}
\newcommand{\proofir}{ProofIR}
\newcommand{\easycrypt}{EasyCrypt}

\makeatletter
\newif\ifarxivversion
\@ifundefined{arxivversion}{\arxivversionfalse}{\arxivversiontrue}
\newif\iflaterresults
\@ifundefined{laterresults}{\laterresultsfalse}{\laterresultstrue}
\makeatother
\newcommand{\arxivonly}[1]{\ifarxivversion#1\fi}
\newcommand{\camerareadyonly}[1]{\ifarxivversion\else#1\fi}

\begin{document}

\ifarxivversion\pagestyle{plain}\fi

\title{\system{}: Towards Automating Formal Cryptographic Proofs}

\author{
\IEEEauthorblockN{
Yiping Ma$^\dagger$,
Yu-Lin Tsai$^\dagger$,
Mayank Rathee$^\dagger$,
Deevashwer Rathee$^\dagger$,\\
Fran\c{c}ois Dupressoir$^\ddagger$,
Pierre-Yves Strub$^\S$,
Raluca Ada Popa$^\dagger$}
\IEEEauthorblockA{
$^\dagger$UC Berkeley \quad
$^\ddagger$University of Bristol \quad
$^\S$PQShield}
}

\hypersetup{
  pdftitle={ShannonProver: Towards Automating Formal Cryptographic Proofs},
  pdfauthor={Yiping Ma, Yu-Lin Tsai, Mayank Rathee, Deevashwer Rathee, Fran\c{c}ois Dupressoir, Pierre-Yves Strub, Raluca Ada Popa}
}

\maketitle
\ifarxivversion\thispagestyle{plain}\fi

\bstctlcite{IEEEexample:BSTcontrol}

\begin{abstract}
Cryptographic proofs are produced at a scale that increasingly exceeds the community’s ability to verify them manually. 
Machine-checked proofs offer a path toward scalable proof verification, 
but they shift the bottleneck to writing the proofs themselves: 
even when the high-level proof plan is known, 
turning it into a proof script requires spelling out every detail the plan leaves implicit, 
which is laborious even for experts. 
This paper presents ShannonProver, an agentic system for automating cryptographic proofs. 
Given a protocol and its security definitions modeled by a cryptographer, ShannonProver decomposes the main theorem into intermediate games and lemmas, and constructs EasyCrypt proof scripts for those lemmas.

We evaluate ShannonProver on a new dataset of lemmas in EasyCrypt. The benchmark spans textbook primitives, deployed standardized protocols, 
as well as recent NIST proposals, and includes expert case studies drawn from a corpus that has not previously been available online.  
On case studies such as ChaChaPoly1305 and MEE-CBC,
ShannonProver completes within hours proof developments that historically took experts weeks to months.
More broadly, this work suggests a path toward accelerating cryptographic research: as agents automate the proof-engineering burden, cryptographers can iterate more quickly on new constructions, obtain machine-checked assurance earlier, and bring protocols from design to deployment faster.
\end{abstract}

\section{Introduction}
\label{sec:introduction}

Cryptographers \textit{``generate more proofs than we carefully verify''}, as
Halevi observed two decades ago~\cite{halevi2005plausible}. This concern is
even more pressing today. Modern cryptographic protocols are increasingly
complex, their proofs are longer and more intricate, and their correctness has
consequences far beyond research papers: they have
become standards~\cite{nistFips202,nistFips197,nistFips203,rfc8446},
libraries~\cite{bernstein2012nacl,libsodiumDocs,opensslGuide,protzenko2020evercrypt},
and security infrastructure~\cite{rfc8446,rfc9001,cohnGordon2017signal,donenfeld2017wireguard}.
Flaws discovered after publication or even standardization
underscore the same point~\cite{almeida2024mlkemArtifact, barbosa2024sphincs}: the community cannot manually scrutinize
cryptographic proofs at the same scale at which it produces them.

Machine-checked proofs offer a way to scale proof verification. Here, a proof is
written as a \emph{proof script}, a sequence of commands called \emph{tactics},
and a proof assistant~\cite{bertot2004coqart,moura2021lean4,swamy2016fstar}
checks the script mechanically.
This work uses \easycrypt{}~\cite{barthe2011workingcryptographer}, a proof assistant designed for game-based cryptographic proofs.
If the check passes, the theorem that the script proves is certified without any expert examining the argument by hand.
However, the bottleneck shifts to producing the proof script in the first place.

Writing such proof scripts is far more exacting than writing traditional pen-and-paper proofs:
a machine-checked proof is not a line-by-line translation of a pen-and-paper proof;
rather, it expands the cryptographer's shorthand of a cryptographic argument (e.g., ``two games are indistinguishable'') into explicit
stateful programs and precisely reasons about the program equivalence.
Historically, developing formal proofs for standardized schemes such as MEE-CBC~\cite{almeida2016meecbc}, CMAC~\cite{baritel2018formal}, and ChaCha20-Poly1305~\cite{almeida2020lastmile} has required several person-weeks to several person-months\footnote{The exact historical effort for the ChaChaPoly development is no longer recoverable; the person-month estimate is based on the size and complexity of the current proof relative to comparable projects with known effort.},
and the effort for recent post-quantum NIST standards such as ML-KEM~\cite{almeida2024mlkem,almeida2024mlkemArtifact} is even larger, involving more than 10 experts for one or two years.

This effort raises a natural question: can experts delegate writing these proof scripts to LLM agents?
Machine-checked proofs make such delegation safe: the agent performs the proof search, which may be heuristic,
while \easycrypt{} performs the proof checking, which is sound.
The agent may hallucinate or make mistakes along the way, but this cannot compromise soundness:
\easycrypt{} rejects every invalid proof attempt, so once the agent produces a script that passes the check,
the theorem is certified.

The most direct approach to this question is to ask the agent to write the entire proof through checker-in-the-loop interaction~\cite{yang2023leandojo,rajaee2025local}:
the agent receives the specification of the scheme and the statement of the main theorem,
submits tactics to \easycrypt{}, receives the resulting goal or error, and repeats until the theorem is proved.
While this approach suffices for simple textbook schemes (e.g., ElGamal), it falls far short of our target:
formal proofs at the scale and complexity of NIST standards and proposals.
We observed that even strong LLMs fail on such proofs.
They sometimes re-prove facts they have already established or drift to the wrong level of abstraction;
as a result, their proofs grow far longer than necessary, or they eventually give up.

Experts do not develop such proofs monolithically.
They proceed in three phases (Fig.~\ref{fig:running-example}):
they first write the \emph{specification}, which formalizes the scheme, its security definitions, and the underlying assumptions (Phase~I),
then decompose the main theorem into intermediate games and lemmas (Phase~II),
and finally prove each lemma with a tactic-level proof script (Phase~III).
Phase~II requires cryptographic insight and high-level planning,
whereas Phase~III is tedious and demands exacting, mechanical proof writing.
The two phases are not sequential: whether a decomposition works is often known only after trying to prove its lemmas.
Experts therefore move back and forth between the two phases until every lemma is proved.
This process is costly: the original MEE-CBC proof took experts a few person-weeks, and the ChaCha20-Poly1305 proof a few person-months.

In this work, we develop \system{}, an agentic system that follows this workflow to automate Phases~II and~III.
It has two components, both built on LLM agents (Fig.~\ref{fig:coordination}): a \emph{decomposer} for Phase~II and a \emph{lemma prover} for Phase~III.
The decomposer works at the level of the whole proof:
it writes the intermediate games and lemma statements, and hands individual lemmas to the lemma prover.
The lemma prover proves each lemma by interacting with \easycrypt{} tactic by tactic.
If it succeeds, it returns a verified proof script;
otherwise, it returns feedback, which the decomposer uses to revise or split the lemma, or to restructure the decomposition.
Because the decomposer can launch the lemma prover on several independent lemmas at once,
\system{} greatly reduces the time needed to test whether a decomposition actually works.
Given only the specification and the main theorem, \system{} constructs complete proofs for MEE-CBC in 2 hours and for ChaCha20-Poly1305 in about 5 hours.

This workflow hinges on the lemma prover: a decomposition can be tested only as quickly and as reliably as its lemmas are proved.
Yet even on a single lemma in Phase~III, the checker-in-the-loop approach can fail when the lemma is difficult or tedious.
These failures have two sources: the agent lacks the context relevant to each step of the proof,
and it explores the proof along a single linear path.
The lemma prover addresses them with two design ideas: a {\em proof-state compiler} and {\em multi-agent tree search}.

\smallskip\noindent\textbf{\ding{182}~Proof-state compiler (\S\ref{sec:proof-state-compiler}).}
The context an agent needs changes as the proof proceeds.
Each tactic transforms the proof state, and with it which {\em proof resources} apply (previously proved lemmas, assumptions, and hypotheses\arxivonly{; App.~\ref{app:proof-resources}})
and how their arguments must be instantiated (left in Fig.~\ref{fig:proof-state-context}).
The agent must derive this context on its own from the relevant files, which is slow;
when it misses something, it often switches to a worse strategy or gets stuck.
Yet this context can be computed deterministically.
Much as a compiler resolves names and types so that programmers need not, the \emph{proof-state compiler} (right in Fig.~\ref{fig:proof-state-context})
analyzes each new state and shows the agent the resources that apply, with their arguments instantiated.
The agent still makes every strategic decision; the compiler takes over the bookkeeping.

\begin{figure}[!bt]
  \centering
  \resizebox{\columnwidth}{!}{\begin{tikzpicture}[
  x=1cm,y=1cm,
  font=\sffamily,
  outerExpert/.style={draw=spcyan!58, fill=softblue, line width=0.55pt,
    rounded corners=2pt},
  outerProver/.style={draw=spamber!58, fill=softamber, line width=0.55pt,
    rounded corners=2pt},
  outerPlanner/.style={draw=spgreen!58, fill=softgreen, line width=0.55pt,
    rounded corners=2pt},
  greenhead/.style={font=\fontsize{5.85}{6.45}\selectfont\fontfamily{lmtt}\selectfont,
    text=spgreen!72!black, anchor=west},
  plannerbadge/.style={draw=spgreen!32, fill=softgreen, rounded corners=1pt,
    line width=0.28pt, font=\fontsize{5.35}{5.85}\selectfont\fontfamily{lmtt}\selectfont,
    text=spgreen!70!black, inner xsep=4.4pt, inner ysep=1.35pt, anchor=east},
  tinyhead/.style={font=\fontsize{5.7}{6.3}\selectfont\sffamily\bfseries,
    text=black!58, anchor=west},
  panelhead/.style={font=\fontsize{5.85}{6.45}\selectfont\fontfamily{lmtt}\selectfont,
    text=spcyan!72!black, anchor=west},
  amberhead/.style={font=\fontsize{5.85}{6.45}\selectfont\fontfamily{lmtt}\selectfont,
    text=codeamber, anchor=west},
  coderow/.style={font=\fontsize{6.05}{6.6}\selectfont\fontfamily{lmtt}\selectfont,
    text=codeblack, anchor=west},
  expertbadge/.style={draw=spcyan!32, fill=softblue, rounded corners=1pt,
    line width=0.28pt, font=\fontsize{5.35}{5.85}\selectfont\fontfamily{lmtt}\selectfont,
    text=spcyan!70!black, inner xsep=4.4pt, inner ysep=1.35pt, anchor=east},
  proverbadge/.style={draw=spamber!32, fill=softamber, rounded corners=1pt,
    line width=0.28pt, font=\fontsize{5.35}{5.85}\selectfont\fontfamily{lmtt}\selectfont,
    text=spamber!78!black, inner xsep=4.4pt, inner ysep=1.35pt, anchor=east},
  panelcut/.style={draw=black!10, line width=0.34pt},
  arrow/.style={->, draw=black!45, line width=0.52pt}
]
\definecolor{spcyan}{RGB}{55,116,145}
\definecolor{spamber}{RGB}{187,124,30}
\definecolor{softblue}{RGB}{238,247,252}
\definecolor{softamber}{RGB}{255,246,230}
\definecolor{spgreen}{RGB}{75,135,80}
\definecolor{softgreen}{RGB}{238,248,240}
\definecolor{panelgray}{RGB}{247,247,244}
\definecolor{panelrule}{RGB}{154,153,145}
\definecolor{codeblack}{RGB}{36,39,40}
\definecolor{codeamber}{RGB}{140,92,25}

\pgfresetboundingbox
\path[use as bounding box] (-0.36,1.29) rectangle (8.92,7.60);

\path[outerExpert] (0.26,5.45) rectangle (8.32,7.47);
\node[panelhead] at (0.58,7.13) {PHASE I \textbullet\ SPECIFICATION};
\node[expertbadge] at (8.00,7.13) {HUMAN EXPERT};
\draw[panelcut] (4.35,6.94) -- (4.35,5.70);
\node[coderow] at (0.58,6.80) {module type IVScheme = \{};
\node[coderow] at (0.82,6.52) {proc keygen()      : key};
\node[coderow] at (0.82,6.24) {proc enc(k,iv,p)   : block list};
\node[coderow] at (0.82,5.96) {proc dec(k,c)      : ...};
\node[coderow] at (0.58,5.68) {\}.};
\node[coderow] at (4.58,6.80) {module CBC(P : PseudoRP) = \{};
\node[coderow] at (4.82,6.52) {proc enc(key,iv,p) = \{ ... \}};
\node[coderow] at (4.58,6.24) {\}.};

\draw[arrow] (4.29,5.45) -- (4.29,5.10);

\path[outerPlanner] (0.26,3.6) rectangle (8.32,5.1);
\node[greenhead] at (0.58,4.80) {PHASE II \textbullet\ DECOMPOSITION};
\node[plannerbadge] at (8.00,4.80) {DECOMPOSER};
\draw[panelcut] (4.35,4.59) -- (4.35,3.70);
\node[coderow] at (0.58,4.46) {lemma cleanup        \&m : Pr... };
\node[coderow] at (0.58,4.18) {lemma CBC\_PRP\_RF     \&m : ...};
\node[coderow] at (0.58,3.90) {lemma CBC\_upto       \&m : ...};
\node[coderow] at (4.58,4.46) {lemma Bound\_by\_PRP\_PRF     \&m : ...};
\node[coderow] at (4.58,4.18) {lemma Bound\_by\_Birthday    \&m : ...};
\node[coderow] at (4.58,3.90) {lemma Conclusion (A)       \&m : ... };
\draw[draw=spamber!70, line width=0.48pt, rounded corners=1pt]
  (0.49,3.76) rectangle (3.38,4.05);

\path[outerProver] (0.26,1.56) rectangle (8.32,3.28);
\node[amberhead] at (0.58,2.97) {PHASE III \textbullet\ LEMMA PROVING};
\node[proverbadge] at (8.00,2.97) {LEMMA PROVER};
\node[coderow] at (0.58,2.63) {proof.};
\node[coderow] at (0.82,2.37) {byequiv=> //=. proc. call (\_: =\{bad,qs\}(DoubleQuery,Compute)).};
\node[coderow] at (0.82,2.09) {by conseq DQ\_Sample\_Compute\_eq. by inline *; auto.};
\node[coderow] at (0.58,1.81) {qed.};

\draw[arrow] (4.29,3.60) -- (4.29,3.28);
\node[font=\fontsize{5.0}{5.5}\selectfont\sffamily\bfseries, text=black!58, anchor=west]
  at (4.45,3.44) {lemmas};
\draw[->, draw=spamber!85, line width=0.6pt, rounded corners=3pt]
  (8.32,2.55) -- (8.60,2.55) -- (8.60,4.55) -- (8.36,4.55);
\node[font=\fontsize{5.0}{5.5}\selectfont\sffamily\bfseries, text=spamber!85!black,
  rotate=90, anchor=center] at (8.76,3.55) {feedback};

\draw[draw=black!45, line width=0.5pt]
  (0.20,5.10) -- (0.06,5.10) -- (0.06,1.56) -- (0.20,1.56);
\node[font=\fontsize{5.6}{6.2}\selectfont\sffamily\bfseries, text=black!62,
  rotate=90, anchor=center] at (-0.13,3.33) {\system{}};

\end{tikzpicture}}
  \caption{The three phases of a formal proof development. A human expert writes the specification (Phase~I), and \system{} automates decomposition (Phase~II) and lemma proving (Phase~III).}
  \label{fig:running-example}
\end{figure}

\begin{figure}[t]
  \centering
  \resizebox{\columnwidth}{!}{%
  \begin{tikzpicture}[
      font=\sffamily\scriptsize,
      seg/.style={draw=black!55, rounded corners=1.5pt, align=center,
                  inner sep=1.5pt, line width=0.45pt, minimum height=0.55cm,
                  anchor=south west, font=\sffamily\tiny},
      job/.style={draw=black!55, rounded corners=1.5pt, align=center,
                  inner sep=1pt, line width=0.45pt, minimum height=0.42cm,
                  anchor=south west, fill=ablours!30, font=\sffamily\tiny},
      hand/.style={-latex, line width=0.5pt, black!65},
      back/.style={-latex, line width=0.5pt, black!65, dashed},
      lane/.style={draw=black!12, line width=0.4pt},
      lbl/.style={font=\sffamily\scriptsize, align=right, anchor=east},
    ]
    \foreach \y in {0.275,-0.73,-1.33} \draw[lane] (1.35,\y) -- (8.75,\y);
    \node[lbl] at (1.3,0.275) {\textbf{Decomposer}\\(Phase II)};
    \node[lbl] at (1.3,-1.03) {\textbf{Lemma prover}\\(Phase III,\\two jobs\\at a time)};

    \node[seg, fill=ablbase!25, minimum width=1.5cm] (o1) at (1.4,0)
      {games G1--G3;\\ lemmas L1--L4};
    \node[seg, fill=ablbase!25, minimum width=1.8cm] (o2) at (2.9,0)
      {prove L3 itself};
    \node[seg, fill=ablbase!25, minimum width=2.2cm] (o3) at (4.7,0)
      {split L4 into L4a, L4b;\\ prove L4b itself};
    \node[seg, fill=ablgrnA!55, minimum width=1.85cm] (o4) at (6.9,0)
      {assemble proof;\\ final EasyCrypt check};

    \node[job, minimum width=1.1cm] (j1) at (3.2,-0.94) {L1 \ding{51}};
    \node[job, minimum width=1.1cm] (j2) at (5.3,-0.94) {L2 \ding{51}};
    \node[job, minimum width=1.8cm] (j3) at (2.9,-1.54) {L4 \ding{55}};
    \node[job, minimum width=1.9cm] (j4) at (5.0,-1.54) {L4a \ding{51}};

    \draw[hand] (3.2,0) -- (3.2,-0.52);
    \draw[back] (4.3,-0.52) -- (4.3,0);
    \draw[hand] (5.3,0) -- (5.3,-0.52);
    \draw[back] (6.4,-0.52) -- (6.4,0);
    \draw[hand] (2.95,0) -- (2.95,-1.12);
    \draw[back] (4.65,-1.12) -- (4.65,0);
    \draw[hand] (5.05,0) -- (5.05,-1.12);
    \draw[back] (6.85,-1.12) -- (6.85,0);

    \draw[-latex, black!45, line width=0.4pt] (1.4,-1.8) -- (8.75,-1.8)
      node[below left=0pt, font=\sffamily\tiny, text=black!60] {time};
  \end{tikzpicture}}
  \caption{Interleaving of Phases II and III in \system{}.
  Solid arrows hand a lemma to the lemma prover; dashed arrows return the
  result, either a proof (\ding{51}) or feedback (\ding{55}). The decomposer
  does not wait for results: while L1 and L4 are being attempted, it proves L3
  itself. When the attempt on L4 fails, the decomposer splits L4 into L4a and
  L4b. Two jobs run at a time here for illustration; the number of concurrent
  jobs is configurable.}
  \label{fig:coordination}
\end{figure}

\smallskip\noindent\textbf{\ding{183}~Multi-agent tree search (\S\ref{sec:tree-policy}).}
A proof rarely proceeds in a straight line: a step that looks reasonable may reveal only much later that it has led to a dead end.
A single agent explores one path at a time, and it often fails to notice when to turn back:
it keeps trying variants of the same failing move instead of returning to where its strategy went wrong.
The lemma prover therefore organizes the search as a tree of checker-accepted proof states (Fig.~\ref{fig:proof-tree}).
Agents can backtrack within their own branches, and a deterministic \emph{orchestrator} oversees all branches:
when an agent stalls, it starts a fresh agent on a different branch, prunes branches that no longer make progress,
and records dead ends so that no agent retries a failed approach at the same state.
Each agent reasons about its own branch; the orchestrator decides where the search should go.

We view this work as a first step toward automating formal proofs for cryptography. Ultimately, we hope \system{} will let experts focus on the critical judgments in formal verification, while automating as much of the remaining work as possible.

\section{Design Overview}
\label{sec:overview}

\subsection{Observations in Agentic Proving}

We observe the failures below in baseline runs, where the agent interacts directly with EasyCrypt (left in Fig.~\ref{fig:proof-state-context}).
They fall into the two groups identified in \S\ref{sec:introduction}:
the agent lacks the context relevant to each step of the proof, and it explores the proof along a single linear path.

\smallskip\noindent\textbf{Missing context.}
At every step, the agent must decide which proof resources apply, which one to use, and how to apply it.
Making this decision requires assembling information scattered across project lemmas, imported theories, module arguments, and even the source code of the proof checker.
The agent can read all of these sources, but it must find and combine the relevant pieces itself, and do so again as the proof state changes.
When it does not, we observe three recurring failure patterns, each at a different point of this decision:
the agent misses a resource that applies, chooses one that does not, or cannot apply the one it has chosen.

\begin{figure}[b]
  \centering
  \resizebox{\linewidth}{!}{\begin{tikzpicture}[
  x=1cm,y=1cm,
  tier/.style={draw=#1, line width=0.55pt, rounded corners=1pt},
  lvl/.style={font=\scriptsize\sffamily, anchor=west, inner sep=0pt},
  tac/.style={font=\scriptsize\ttfamily, text=black!62, anchor=west, inner sep=0pt},
  eff/.style={font=\scriptsize\sffamily, align=right, anchor=east, inner sep=0pt},
  down/.style={font=\scriptsize\ttfamily, text=black!72, anchor=east, inner sep=0pt},
  axis/.style={font=\scriptsize\sffamily, text=black!65}
]
\definecolor{spgreen}{RGB}{75,135,80}
\definecolor{spamber}{RGB}{187,124,30}
\definecolor{spred}{RGB}{175,72,60}
\definecolor{splightblue}{RGB}{82,151,190}
\definecolor{softgreen}{RGB}{238,248,240}
\definecolor{softamber}{RGB}{255,246,230}
\definecolor{softred}{RGB}{253,239,237}
\definecolor{softblue}{RGB}{238,247,252}

\def\step{0.9}
\def\h{0.62}
\def\xz{-2.4}
\def\dx{0.45}
\def\xr{5.3}
\def\xt{0.85}

\node[axis] at (1.45,0.34) {semantic / proof-structure level};
\draw[black!28, line width=0.55pt] (\xz,0.16) -- (\xr,0.16);

\foreach \i/\name/\moves/\effect/\col/\fill in {
  0/{Pr expression algebra}/{rewrite, congr}/{program structure\\preserved}/spgreen/softgreen,
  1/{pRHL program equivalence}/{sim, conseq}/{program structure\\preserved}/spgreen/softgreen,
  2/{Module-level call}/{call (\_: Inv)}/{call abstraction\\preserved}/spgreen/softgreen,
  3/{Procedure instructions}/{wp, rnd, sp}/{procedure body\\exposed}/spamber/softamber,
  4/{All-modules expansion}/{wp, rnd, sp}/{all abstractions\\expanded}/spred/softred,
  5/{Atomic / SMT discharge}/{smt(), auto}/{atomic goal\\discharged}/splightblue/softblue} {
  \pgfmathsetmacro{\y}{-\i*\step}
  \pgfmathsetmacro{\xl}{\xz-\i*\dx}
  \path[tier=\col, fill=\fill] (\xl,\y) rectangle (\xr,\y-\h);
  \node[lvl] at (\xl+0.18,\y-\h/2) {\name};
  \node[tac] at (\xt,\y-\h/2) {\moves};
  \node[eff, text=\col] at (\xr-0.15,\y-\h/2) {\effect};
}

\foreach \i/\d in {0/byequiv, 1/proc, 2/{inline f}, 3/{inline *}, 4/skip} {
  \pgfmathsetmacro{\ya}{-\i*\step-\h}
  \pgfmathsetmacro{\yb}{-(\i+1)*\step}
  \pgfmathsetmacro{\xa}{\xz-\i*\dx+0.3}
  \draw[-latex, line width=0.6pt, draw=black!50] (\xa,\ya) -- (\xa,\yb);
  \node[down] at (\xa-0.12,{(\ya+\yb)/2}) {\d};
}

\draw[->, line width=0.8pt, draw=black!45] (-5.3,0.35) -- (-5.3,-5.15);
\node[axis, rotate=90] at (-5.6,-2.4) {lowering};

\draw[black!28, line width=0.55pt] (-5.0,-5.3) -- (\xr,-5.3);
\node[axis] at (0.15,-5.48) {syntactic / verification-condition level};
\end{tikzpicture}}
  \caption{The tactic pyramid: abstraction levels in an EasyCrypt proof, from probability expressions down to SMT goals.
  Each level lists the tactics used within it; each arrow between two levels is labeled with the tactic that moves the proof down.
  Lowering exposes program detail but can invalidate resources that apply only at higher levels.}
  \label{fig:tactic-pyramid}
\end{figure}

\noindent\ding{182}~\emph{Unconscious lowering.}
A proof often moves across several abstraction levels (Fig.~\ref{fig:tactic-pyramid}):
from probability algebra and pRHL equivalences, to module or oracle calls, procedure bodies,
and finally SMT side conditions.
Higher levels let the prover reuse facts already proved about whole games, modules, or oracle behaviors.
Human experts descend to a lower level only when the proof really needs more concrete program structure; such a decision is guided by experience and by the intuition they gained from lemma decomposition (Phase~II).
However, an LLM agent does not always make an accurate judgment about when to descend;
the agent can mistakenly interpret an unhelpful goal change as real proof progress. In the baseline runs, a tactic such as \texttt{inline*} is very tempting because it is easy to write and immediately produces a different-looking goal. What the agent does not see is that this change may cut off resources that are usable at the higher level: a lemma stated about a procedure call, such as an equivalence between two oracle implementations, applies only while that call is still visible in the goal; once the call is inlined, the lemma no longer matches (App.~\ref{app:failures-and-standouts}). The agent must then either reprove from scratch what the lemma would have established, or keep pursuing the original plan that relied on the lemma but can no longer succeed from this state. The failure is not that the agent chose to descend; descent is often necessary. The failure is that the interface gives no state-aware information when a locally productive-looking move is globally destructive.

\noindent\ding{183}~\emph{Semantic mislocalization.}
An agent can choose a resource that does not apply.
This is not simply a lack of information: the proof context may contain the right clues, but
the agent can be pulled toward the wrong resource by names that appear in the goal
or by the literal wording of an error message. For example, in one run, a goal
mentioning \texttt{IFinRO} and \texttt{IndRO} led the agent to treat these names
as evidence for an eager/lazy random-oracle lemma and to spend the rest of the
run trying to fit \texttt{pr\_RO\_FinRO\_D}, which is unrelated to the proof (App.~\ref{app:failures-and-standouts}).

\noindent\ding{184}~\emph{Agent flinch.}
An agent can choose the right resource and still fail to apply it.
Applying a lemma means turning the choice into a command that EasyCrypt accepts:
looking up the lemma's exact statement, supplying its module arguments, matching types, and finding the right tactic form.
When this information is available only through trial and error with the checker,
the agent often abandons the plan after a few rejected attempts and switches to a tactic that is easier to write,
even if it leads to a worse proof state (App.~\ref{app:failures-and-standouts}).
The failure is not the absence of a proof idea, but the cost of carrying it out.

\smallskip\noindent\textbf{Exploring a single path.}
Even with the right context at every step, a tactic that EasyCrypt accepts can lead the proof into a dead end that becomes visible only many steps later.
The agent must therefore also decide when to abandon its current strategy and where to resume.
Even an agent that can return to an earlier state often does not notice when it should:
it stays at a state and keeps submitting variants of the same failing move,
or keeps repairing a branch whose strategy no longer works,
instead of returning to the point where the strategy went wrong.
And when it does backtrack, it pursues one strategy at a time, so several plausible strategies can only be tried one after another.
These failures do not come from a single wrong step, but from the lack of any view of the search as a whole.

\subsection{Design Ideas}

\smallskip\noindent\textbf{Proof-state compiler.}
The three failure patterns point to an interface problem:
the agent must reconstruct, through ad hoc queries such as \texttt{Grep} and \texttt{Read}, proof context that is already determined by the current proof state and the project.
\system{} instead computes this context deterministically with a \emph{proof-state compiler} (\S\ref{sec:proof-state-compiler}, Fig.~\ref{fig:proof-state-context}).
At each proof state, the compiler takes as input the checker's output, the proof history, and the project context, including source files, imported theories, module definitions, and available lemmas.
Its output addresses the three failure patterns in turn:
it shows which resources still apply at the current state, so that the agent does not lower past them;
it matches resources against the goal by structure rather than by name;
and once the agent selects a resource, it fills in the arguments and premises that the current state determines, so that the agent can carry its plan through.
These outputs are facts about the current proof state, not a proof strategy: the agent still decides which proof operation and resource to use next.
The compiler's analyses follow classical compiler design, which builds an intermediate representation, resolves names and types, tracks which facts remain valid, and reports diagnostics~\cite{aho2006compilers};
Fig.~\ref{fig:route-ladder} shows its three layers.

\smallskip\noindent\textbf{Multi-agent tree search.}
Neither failure of single-path exploration can be fixed inside one agent:
an agent that does not notice when to turn back cannot be relied on to turn back,
and a single agent can pursue only one strategy at a time.
\system{} therefore organizes the search as a tree of checker-accepted proof states (\S\ref{sec:tree-policy}, Fig.~\ref{fig:proof-tree}).
Each agent can backtrack within its own branch, which handles the wrong turns it recognizes.
A deterministic orchestrator watches the progress of all branches and handles the rest:
when an agent stalls, it starts a fresh agent on a different branch,
and because several branches run at once, plausible strategies are tried in parallel rather than one after another.
It also prunes branches that fall behind and records the attempts that failed.

\section{Proof-State Compiler}
\label{sec:proof-state-compiler}

Writing an EasyCrypt proof requires two kinds of work. First, the prover makes a strategic choice: it selects a proof operation and the proof resources with which to carry out that operation. By a \emph{proof operation} we mean a kind of proof step rather than one specific command: for example, using a lemma about a procedure call (\texttt{call}, \texttt{ecall}), rewriting with an equation (\texttt{rewrite}), or turning a probability statement into a program equivalence (\texttt{byequiv}). The same operation can be written with different tactics and tactic forms. A \emph{proof resource} is something the proof can use instead of reasoning from scratch, such as a lemma already proved in the project or in a library, an axiom of the specification, or a hypothesis in the current proof. Which resources can be used changes as the proof proceeds\arxivonly{ (App.~\ref{app:proof-resources})}. Second, the prover must realize that choice as an EasyCrypt command that is valid at the exact current proof state. This requires resolving names and modules, supplying the appropriate arguments, and using the correct tactic form.

The first kind of work is \emph{strategic}: it requires proof insight, that is, semantic reasoning about what the programs and statements mean. The second is \emph{mechanical}: it is often determined by the syntactic form of the current goal and of the selected resource, and therefore changes with every proof state. Repeatedly working on such details through trial and error consumes reasoning budget on work that can instead be performed deterministically. More importantly, we observed that these low-level failures can cause an agent to abandon an otherwise correct high-level plan and fall back to a worse proof strategy, as in the agent flinch of Section~\ref{sec:overview}.

We therefore separate strategic choices from their mechanical realization. 
Concretely, the agent chooses the proof operation, the proof resource, and any other arguments that require proof insight, such as an invariant. 
The proof-state compiler then \emph{realizes} the agent's intended step: it keeps the proof operation and resource that the agent chose, fills in what the current state determines, and uses the tactic form that EasyCrypt expects; everything else is left to the agent.
In particular, the compiler never replaces the agent's choices. Changing the proof strategy is also the agent's decision; Section~\ref{sec:tree-policy} describes the backtracking that supports it.

\smallskip
\noindent
{\em Running example.}
We use a simple losslessness proof throughout this section to illustrate our design (Fig.~\ref{fig:proof-state-context}). The goal is to prove that the procedure \texttt{D2(O).distinguish}, shown below in simplified form, is lossless:
\begin{verbatim}
proc D2(O).distinguish() = {
  D2(O).O.init();
  b <@ Adv_MAC_to_F(A, D2(O).O).guess();
  return b;
}
\end{verbatim}
Elsewhere in the project, the following lemma has been proved:
\begin{verbatim}
lemma Alossless_F (X <: PRF_Oracles) :
  islossless X.f =>
  islossless Adv_MAC_to_F(A, X).guess.
\end{verbatim}
After opening the procedure with \texttt{proc}, the agent correctly decides to use the proof operation \texttt{call} with the resource \texttt{Alossless\_F}, but submits \texttt{call (Alossless\_F \_ \_).}
EasyCrypt rejects the command because it cannot infer the module argument. The intended step should be realized as
\begin{verbatim}
    call (Alossless_F (<: D2(O).O) _).
\end{verbatim}
 
In the agent's attempt, the first hole is a module argument, which the current state determines. The second hole is a premise of the lemma. The compiler fills it only if the current proof context already has a hypothesis stating that premise; otherwise the premise is a new proof obligation that requires semantic reasoning, and it is left to the agent.

\begin{figure*}[t]
  \centering
  \resizebox{\textwidth}{!}{\colorlet{figecgray}{black!40}
\begin{tikzpicture}[
  font=\sffamily\footnotesize,
  role/.style={draw=#1, fill=#1!16, rounded corners=2pt, line width=0.6pt,
    minimum height=0.82cm, align=center, inner sep=2pt,
    font=\sffamily\footnotesize\bfseries, text=black!80},
  files/.style={draw=black!32, dashed, rounded corners=2pt, fill=black!3,
    minimum height=0.5cm, align=center, inner sep=2pt,
    font=\sffamily\scriptsize, text=black!65},
  flow/.style={-latex, line width=0.55pt, draw=black!55},
  lab/.style={font=\sffamily\scriptsize, text=black!60, inner sep=1.5pt},
  side/.style={font=\sffamily\footnotesize\bfseries, anchor=west, inner sep=0pt},
  paneltag/.style={font=\sffamily\scriptsize, text=black!55, anchor=west, inner sep=0pt},
  sec/.style={font=\sffamily\scriptsize\bfseries, text=black!72, anchor=north west, inner sep=0pt},
  ref/.style={font=\sffamily\scriptsize, text=black!45, anchor=north east, inner sep=0pt},
  code/.style={font=\ttfamily\scriptsize, text=black!88, anchor=north west, inner sep=0pt},
  note/.style={font=\sffamily\scriptsize, text=black!62, anchor=north west, inner sep=0pt},
]

\node[side, text=ablbase!55!black] at (0,0) {Without the compiler};
\node[side, text=ablours!55!black] at (9.3,0) {With the proof-state compiler};
\draw[black!15, dashed, line width=0.5pt] (8.9,0.2) -- (8.9,-6.9);

\node[role=ablbase, minimum width=1.7cm] (agentL) at (0.85,-0.9) {LLM agent};
\node[role=figecgray, minimum width=1.7cm] (ecL) at (7.65,-0.9) {EasyCrypt};
\draw[flow] ([yshift=0.14cm]agentL.east) -- node[lab, above] {tactic} ([yshift=0.14cm]ecL.west);
\draw[flow] ([yshift=-0.14cm]ecL.west) -- node[lab, below] {goal or error} ([yshift=-0.14cm]agentL.east);
\node[files, minimum width=1.7cm] (filesL) at (0.85,-2.05) {project files};
\draw[flow] (filesL.north) -- node[lab, right=3pt] {agent searches and reads} (agentL.south);

\node[role=ablbase, minimum width=1.7cm] (agentR) at (10.15,-0.9) {LLM agent};
\node[role=ablours, minimum width=2.7cm] (compR) at (13.55,-0.9) {Proof-state\\[-1pt]compiler};
\node[role=figecgray, minimum width=1.7cm] (ecR) at (16.95,-0.9) {EasyCrypt};
\draw[flow] ([yshift=0.14cm]agentR.east) -- node[lab, above] {attempt} ([yshift=0.14cm]compR.west);
\draw[flow] ([yshift=-0.14cm]compR.west) -- node[lab, below] {view} ([yshift=-0.14cm]agentR.east);
\draw[flow] ([yshift=0.14cm]compR.east) -- node[lab, above] {tactic} ([yshift=0.14cm]ecR.west);
\draw[flow] ([yshift=-0.14cm]ecR.west) -- node[lab, below] {goal} ([yshift=-0.14cm]compR.east);
\node[files, minimum width=2.7cm] (filesR) at (13.55,-2.05) {project files};
\draw[flow] (filesR.north) -- node[lab, right=3pt] {compiler input} (compR.south);

\node[paneltag] at (0,-2.72) {What the agent sees};
\node[paneltag] at (9.3,-2.72) {What the agent sees};
\draw[draw=ablbase, line width=0.6pt, rounded corners=3pt, fill=ablbase!5] (0,-2.95) rectangle (8.5,-6.85);
\draw[draw=ablours, line width=0.6pt, rounded corners=3pt, fill=ablours!6] (9.3,-2.95) rectangle (17.8,-6.85);

\node[sec] at (0.3,-3.2) {Goal after \texttt{proc}};
\node[code] at (0.45,-3.55) {D2(O).O.init();};
\node[code] at (0.45,-3.87) {b <@ Adv\_MAC\_to\_F(A, D2(O).O).guess();};
\node[sec] at (0.3,-4.35) {The agent's attempt};
\node[code] at (0.45,-4.7) {call (Alossless\_F \_ \_).};
\node[note] at (0.45,-5.02) {\ding{55}\ \ rejected: EasyCrypt cannot infer the module argument};
\node[sec] at (0.3,-5.82) {What the agent must do next};
\node[note] at (0.45,-6.17) {find the module argument, the premise, and the tactic form};
\node[note] at (0.45,-6.49) {by searching the project files and retrying};

\node[sec] at (9.6,-3.2) {Goal after \texttt{proc}};
\node[ref] at (17.5,-3.2) {\S\ref{subsec:projection}};
\node[code] at (9.75,-3.55) {D2(O).O.init();};
\node[code] at (9.75,-3.87) {b <@ Adv\_MAC\_to\_F(A, D2(O).O).guess();};
\node[sec] at (9.6,-4.35) {Applicable resources};
\node[ref] at (17.5,-4.35) {\S\ref{subsec:action}};
\node[code] at (9.75,-4.7) {Alossless\_F};
\node[note] at (11.35,-4.7) {matches the call to \texttt{guess}};
\node[code] at (9.75,-5.02) {\ \ X := D2(O).O};
\node[code] at (9.75,-5.34) {\ \ premise: islossless D2(O).O.f};
\node[sec] at (9.6,-5.82) {Realized tactic, after the agent selects \texttt{Alossless\_F}};
\node[ref] at (17.5,-5.82) {\S\ref{subsec:liveness}};
\node[code] at (9.75,-6.17) {call (Alossless\_F (<: D2(O).O) \_).};
\node[note] at (14.25,-6.17) {\ding{51}\ \ submitted to EasyCrypt};
\node[code, text=black!60] at (9.75,-6.49) {\ \ new subgoal: islossless D2(O).O.f};
\end{tikzpicture}}
  \caption{One step of the running example without (left) and with (right) the proof-state compiler.
  Left: the agent's attempt \texttt{call (Alossless\_F \_ \_).} is rejected because EasyCrypt cannot infer the module argument, and the agent must search the project files to recover it.
  Right: the compiler shows the resources that apply at the current state, with their arguments instantiated (\S\ref{subsec:action}); once the agent selects \texttt{Alossless\_F}, the compiler realizes the tactic and submits it (\S\ref{subsec:liveness}).
  The agent chooses the resource; the compiler fills in only what the state determines and leaves the lemma's premise to the agent as a new subgoal.}
  \label{fig:proof-state-context}
\end{figure*}

\smallskip
\noindent
\textbf{Design overview.}
Our compiler is organized into three layers (Fig.~\ref{fig:route-ladder}).
The first layer, state projection (\S\ref{subsec:projection}), determines which proof state the agent is working on; all later analysis refers to this state.
The second layer, ProofIR (\S\ref{subsec:proofir}), parses the current goal into a typed representation, together with the proof operation and resource of the agent's attempt, if there is one.
The third layer, resource and binding analysis, uses these results in one of two ways, depending on whether the agent has already selected a resource.
If it has not, \emph{resource exposure} (\S\ref{subsec:action}) finds the resources that apply at the current state and shows them to the agent.
If it has, \emph{binding analysis} (\S\ref{subsec:liveness}) realizes the agent's intended step.

\begin{figure}[t]
  \centering
  \resizebox{\columnwidth}{!}{\begin{tikzpicture}[
  x=1cm,y=1cm,
  font=\sffamily,
  stage/.style 2 args={draw=#1, fill=#2, line width=0.6pt,
    rounded corners=2.2pt},
  stagetitle/.style={font=\fontsize{7.4}{8.0}\selectfont\sffamily\bfseries,
    text=black!86, anchor=west},
  stagebody/.style={font=\fontsize{6.05}{6.65}\selectfont\sffamily,
    text=black!68, anchor=west},
  output/.style={font=\fontsize{5.95}{6.5}\selectfont\sffamily,
    text=black!72, anchor=west},
  number/.style={font=\fontsize{7.2}{7.8}\selectfont\sffamily\bfseries},
  arrow/.style={->, draw=black!42, line width=0.6pt},
  casehead/.style={font=\fontsize{6.55}{7.1}\selectfont\sffamily\bfseries,
    anchor=west},
  casenote/.style={font=\fontsize{5.45}{6.0}\selectfont\sffamily,
    text=black!58, anchor=west},
  casebody/.style={font=\fontsize{5.85}{6.4}\selectfont\sffamily,
    text=black!70, anchor=west, align=left},
  caseoutput/.style={font=\fontsize{5.75}{6.3}\selectfont\sffamily,
    text=black!72, anchor=west, align=left},
  flowlabel/.style={font=\fontsize{5.4}{5.95}\selectfont\sffamily,
    text=black!58, anchor=west}
]

\definecolor{spteal}{HTML}{12978C}
\definecolor{spviolet}{HTML}{7A63B0}
\definecolor{spamber}{RGB}{187,124,30}
\definecolor{softamber}{RGB}{255,246,230}
\definecolor{softgray}{RGB}{248,248,246}

\newcommand{\compilerstage}[6]{%
  \path[stage={#6!70}{#6!6}] (0,#1-0.94) rectangle (8.48,#1);
  \path[fill=#6!16] (0,#1-0.94) rectangle (0.72,#1);
  \draw[#6!70, line width=0.45pt] (0.72,#1-0.94) -- (0.72,#1);
  \node[number, text=#6!66!black] at (0.36,#1-0.47) {#2};
  \node[stagetitle] at (0.94,#1-0.18) {#3};
  \node[stagebody] at (0.94,#1-0.43) {#4};
  \draw[black!10, line width=0.32pt] (0.94,#1-0.59) -- (8.18,#1-0.59);
  \node[output] at (0.94,#1-0.77) {\textbf{output:} #5};
}

\path[draw=black!24, fill=softgray, line width=0.55pt, rounded corners=2.2pt]
  (0,9.60) rectangle (8.48,10.40);
\node[font=\fontsize{6.3}{6.9}\selectfont\sffamily\bfseries,
  text=black!72, anchor=west] at (0.24,10.19) {INPUT};
\node[font=\fontsize{6.15}{6.7}\selectfont\sffamily,
  text=black!72, anchor=west] at (1.08,10.19)
  {current EasyCrypt state and project files};
\node[font=\fontsize{6.15}{6.7}\selectfont\sffamily,
  text=black!72, anchor=west] at (1.08,9.83)
  {optional agent choice: proof operation + resource + other arguments};

\draw[arrow] (4.24,9.60) -- (4.24,9.42);

\compilerstage{9.38}{1}{State projection}
  {Attach goals, errors, and derived facts to the current state.}
  {a view of the current state; facts from other states are not shown}{spteal}

\draw[arrow] (4.24,8.44) -- (4.24,8.26);

\compilerstage{8.22}{2}{ProofIR}
  {Parse the goal and the agent's attempt into a typed representation.}
  {typed goal and program; the role of each missing argument}{spviolet}

\draw[arrow] (4.24,7.28) -- (4.24,7.10);

\path[stage={spamber!70}{softamber}] (0,3.40) rectangle (8.48,7.06);
\path[fill=spamber!16] (0,3.40) rectangle (0.72,7.06);
\draw[spamber!70, line width=0.45pt] (0.72,3.40) -- (0.72,7.06);
\node[number, text=spamber!66!black] at (0.36,5.23) {3};
\node[stagetitle] at (0.94,6.88) {Resource and binding analysis};
\node[stagebody] at (0.94,6.62)
  {Depends on whether the agent has selected a resource.};

\path[draw=spamber!55, fill=white, line width=0.5pt, rounded corners=1.8pt]
  (0.94,5.28) rectangle (8.18,6.40);
\path[fill=spamber!10] (0.94,5.28) rectangle (3.54,6.40);
\draw[spamber!35, line width=0.35pt] (3.54,5.28) -- (3.54,6.40);
\node[casehead, text=spamber!68!black] at (1.12,6.14) {Resource exposure};
\node[casenote] at (1.12,5.82) {no resource selected};
\node[casebody] at (3.66,6.12)
  {Match available resources to the\\current goal or program.};
\draw[black!10, line width=0.32pt] (3.66,5.84) -- (7.98,5.84);
\node[caseoutput] at (3.66,5.54)
  {\textbf{output:} all matching resources,\\instantiated at the current state};

\draw[arrow] (4.24,5.28) -- (4.24,4.92);
\node[flowlabel] at (4.48,5.10) {agent selects a resource};

\path[draw=spamber!55, fill=white, line width=0.5pt, rounded corners=1.8pt]
  (0.94,3.58) rectangle (8.18,4.92);
\path[fill=spamber!22] (0.94,3.58) rectangle (3.54,4.92);
\draw[spamber!35, line width=0.35pt] (3.54,3.58) -- (3.54,4.92);
\node[casehead, text=spamber!68!black] at (1.12,4.66) {Binding analysis};
\node[casenote] at (1.12,4.34) {resource selected};
\node[casebody] at (3.66,4.64)
  {Fill arguments and premises fixed by\\the state; EasyCrypt elaborates.};
\draw[black!10, line width=0.32pt] (3.66,4.34) -- (7.98,4.34);
\node[caseoutput] at (3.66,3.98)
  {\textbf{output:} one realized tactic (submitted),\\or an explanation or a list of completions};

\pgfresetboundingbox
\path[use as bounding box] (-0.04,3.34) rectangle (8.52,10.44);
\end{tikzpicture}}
  \caption{The three compiler layers. The third layer either shows the agent
  the resources that match the current state or realizes a resource that the
  agent has already selected.}
  \label{fig:route-ladder}
\end{figure}

\subsection{State Projection: Where am I in the proof?}
\label{subsec:projection}

Writing a formal proof for a lemma is often highly local. 
At each step, the information needed to construct the next tactic sequence is often specific to the current proof state. 
An accepted tactic advances the proof to a new state, a rejected tactic reports an error while leaving the state unchanged, 
and an undo restores an earlier state. 
Consequently, the interaction transcript between the agent and EasyCrypt interleaves goals,
error messages, and analyses associated with different proof states.

We do not ask the agent to keep the history. Instead, the compiler maintains a state-centered view consisting of the current goal together with information derived specifically for that state. When the proof advances or returns to an earlier state, the view is recomputed for that state.  

This design serves two purposes. First, it removes bookkeeping from the agent: the agent does not need to determine which parts of a long transcript are still relevant. Second, it prevents stale information from silently influencing later proof decisions. An error produced before an undo, for example, should not become a general statement that a tactic or resource is unusable.

Long-term reasoning is not the compiler's job. The agent keeps its overall plan and the proof strategy it is following, 
and the proof tree of Section~\ref{sec:tree-policy} records which alternatives have already failed.
State projection only ensures that the proof facts presented by the compiler describe where the agent is \emph{now}.

\smallskip
\noindent
{\em Running example.}
After the agent runs \texttt{proc}, the goal shows the body of \texttt{D2(O).distinguish}. The rejected attempt \texttt{call (Alossless\_F \_ \_).} does not change the proof state, so its error is attached to the state after \texttt{proc}, and the compiler analyzes that same state. If the agent later undoes \texttt{proc} or moves to a different state, this error is no longer shown as if it described the new state.

\subsection{ProofIR: How to interpret the state?}
\label{subsec:proofir}

Once the current state is fixed, the compiler must determine what part of the agent's proof attempt is already a strategic choice and what part is an incomplete mechanical detail. Raw tactic syntax does not make this distinction explicit. An underscore in an EasyCrypt term, for example, may stand for a module parameter, a value argument, or an unresolved proof premise. Treating all of these simply as ``missing syntax'' would give the compiler no principled way to decide what it may fill in.

ProofIR makes these roles explicit. For each proof attempt of the agent, it records three kinds of information:

\noindent\ding{182}~The current goal and relevant program structure.

\noindent\ding{183}~The proof operation and proof resource selected by the agent.

\noindent\ding{184}~The typed parameters of that resource, including which arguments were supplied and which remain unresolved.

If the agent has not submitted an attempt, ProofIR records only the first kind of information, which resource exposure (\S\ref{subsec:action}) uses.

ProofIR turns an informal failure such as ``this command did not elaborate'' into a precise state-dependent question about the proof operation and resource that the agent has already selected.

\smallskip
\noindent
{\em Running example.}
The ProofIR layer parses the agent's attempt \texttt{call (Alossless\_F \_ \_).} It records that the requested proof operation is \texttt{call} and the selected resource is a lemma \texttt{Alossless\_F}. It parses the lemma as 
\begin{verbatim}
module param  : X
module type   : PRF_Oracles
premise       : islossless X.f
conclusion    : islossless Adv_MAC_to_F(A, X).guess
\end{verbatim}
In this representation, the role of each missing argument in the agent's attempt is explicit: 
the first hole asks for a module instantiation; the second asks for a proof of a losslessness premise. 
This layer only parses: it does not fill either hole or assert that the lemma applies. It makes explicit what binding analysis (\S\ref{subsec:liveness}) must relate.

\subsection{Resource Exposure: What can I use?}
\label{subsec:action}

Resource exposure handles the case in which the agent has not yet selected a resource.
This addresses unconscious lowering and semantic mislocalization (\S\ref{sec:overview}), in which the agent misses a resource that applies or chooses one that does not.
For example, after a procedure body is opened, the goal shows a procedure call,
but it does not show which existing lemmas can be used at that call.

The compiler therefore finds resources that are {\em structurally} applicable.
This matching is syntactic: the compiler compares the form of the current goal with the form of each resource's statement,
rather than relying on the similarity of names or text. Because the matching is syntactic, we do it deterministically instead of using an LLM.
It proceeds in three steps.

\noindent\ding{182}~The compiler finds the objects in the current goal against which resources can be matched, such as procedure calls.
In the running example, after the proof operation \texttt{proc} and before the agent submits any \texttt{call}, the goal contains the two statements of the procedure body shown above. The last statement is the procedure call \texttt{Adv\_MAC\_to\_F(A, D2(O).O).guess}, which is where a lemma about a call can be applied with \texttt{call}.

\noindent\ding{183}~The compiler finds the resources available in the current file and its imports\arxivonly{ (App.~\ref{app:proof-resources})} whose statements match that call:
\begin{verbatim}
resource     :   Adv_MAC_to_F(A, X       ).guess
current call :   Adv_MAC_to_F(A, D2(O).O ).guess
\end{verbatim}
It checks that the procedure names are the same (\texttt{guess}), that the enclosing module application is the same (\texttt{Adv\_MAC\_to\_F}), that the module arguments already present are the same (\texttt{A}), and that the remaining module parameter \texttt{X} can be instantiated with the concrete module \texttt{D2(O).O} from the current state.

\noindent\ding{184}~The compiler shows the agent each matching resource, with its statement instantiated at the current state, including the premises that remain to be proved. Here, the agent is shown \texttt{Alossless\_F} with
\begin{verbatim}
premise   : islossless D2(O).O.f
conclusion: islossless
              Adv_MAC_to_F(A, D2(O).O).guess
\end{verbatim}
If several resources match, the compiler shows all of them, without ranking them, and the agent chooses.
Once the agent selects a resource, binding analysis (\S\ref{subsec:liveness}) completes the tactic.

\subsection{Binding Analysis: How should I execute?}
\label{subsec:liveness}

Binding analysis handles the case in which the agent has selected a proof operation and a resource, either on its own, as in the running example, or from those shown by resource exposure. It realizes the agent's intended step.
This addresses agent flinch (\S\ref{sec:overview}): the agent has chosen the right resource but cannot turn the choice into a command that EasyCrypt accepts.
We allow the compiler to fill an argument only when that argument is determined by the selected proof operation, the selected resource, and the current proof state.
The analysis proceeds in three steps.

\noindent\ding{182}~It determines where the attempted proof operation applies in the current state and checks whether the selected resource applies there. This applicability is inherently state-dependent: a lemma about a procedure call, for example, can be used only while the corresponding call is still visible in the goal; after the procedure is inlined or the proof moves past that call, the same lemma may no longer apply (the unconscious lowering of \S\ref{sec:overview}; Fig.~\ref{fig:tactic-pyramid}).

\noindent\ding{183}~When the selected resource does apply, the compiler uses its typed structure to determine missing arguments. Module parameters, procedure parameters, or other bindings can often be read off by comparing the resource with the program object to which it is being applied. A premise of the resource is filled only if a hypothesis in the current proof context states exactly that premise. This is mechanical: the agent has chosen the resource, and the hypothesis is determined by the premise. Otherwise the premise is left as a hole, because it is a new proof obligation. Finally, the compiler writes the instantiated resource in the form that the proof operation expects (for example, marking a module argument with \texttt{<:}).

\noindent\ding{184}~The compiler asks EasyCrypt to elaborate the instantiated term. If elaboration fails, the proposed binding is discarded; the compiler does not substitute a different resource.
If this yields exactly one tactic, the compiler submits it to EasyCrypt. Otherwise, it tells the agent why no tactic was produced, for example that the selected resource does not apply at the current state; if several completions are possible, it lists them and lets the agent choose.

\smallskip
\noindent
{\em Running example.}
As in \S\ref{subsec:action}, comparing the call \texttt{Adv\_MAC\_to\_F(A, D2(O).O).guess} with the statement of the selected lemma determines the module argument uniquely as \texttt{X := D2(O).O}, which leaves the premise \texttt{islossless D2(O).O.f}.
If the current proof context has a hypothesis \texttt{Ofll : islossless D2(O).O.f}, for example an assumption of the lemma being proved, the compiler uses it, and the realized tactic is
\begin{verbatim}
    call (Alossless_F (<: D2(O).O) Ofll).
\end{verbatim}
Otherwise, the premise is a new proof obligation. The compiler leaves it as a hole, and the realized tactic is
\begin{verbatim}
    call (Alossless_F (<: D2(O).O) _).
\end{verbatim}
EasyCrypt then turns the hole into a new subgoal, \texttt{islossless D2(O).O.f}, which the agent proves.

\section{Tree-Based Proof Orchestration}
\label{sec:tree-policy}

Section~\ref{sec:overview} observed that a single agent explores the proof along one path and often does not notice when it should turn back.
\system{} addresses this with a tree-based search explored at two levels.
Within a single agent, the agent can backtrack and re-explore an earlier
state when it discovers that the current strategy is unproductive. Across agents,
an orchestrator manages several branches of the proof tree: it decides when
to start a new agent on a fresh branch, when to stop an unpromising one, and how failures found
on one branch should guide the others. We first introduce the basic tree
concepts (\S\ref{subsec:proof-tree}).

\subsection{Proof Tree}
\label{subsec:proof-tree}

A \emph{node} represents a proof state. Working on a proof means moving progressively between nodes. An \emph{edge} is an accepted transition between states: the agent submits a tactic, the checker accepts it, and the proof state moves from the current node to a child node. A rejected tactic leaves the state
unchanged and creates no edge. A \emph{path} is the sequence of nodes from the root to the current node, representing proof attempts. An \emph{agent} is a prover instance that extends a single branch of the tree, advancing the proof one node at a time.

The search space is a tree rather than a single line because the same state can
be continued in more than one way. A node gains another child in two ways.
First, an agent may \emph{backtrack} to that node and try a different strategy;
if the new tactic is accepted, it leads to a different child state
(\S\ref{subsec:backtracking}). Second, the orchestrator may \emph{spawn} another
agent at the same shared state, so that a different strategy can be explored
without waiting for the current branch to finish (\S\ref{subsec:scheduling}).
The resulting branches are \emph{siblings} below a common node. A branch ends at
a \emph{leaf}: either a closed proof (\texttt{qed.}) or a state that is
abandoned because it is a dead end or is pruned by the orchestrator
(\S\ref{subsec:scheduling}).
\arxivonly{If every one of these nodes could branch, the tree would grow exponentially with the length of the proof.
In practice, only a few states are worth branching from; \S\ref{subsec:scheduling} describes how the orchestrator chooses them.}

\subsection{Backtracking within an Agent}
\label{subsec:backtracking}

At each node, the agent reads the output of the proof-state compiler
(Section~\ref{sec:proof-state-compiler}), which gives it a structured view of the current proof state. Based on this view, the agent extends its branch one tactic at a time, then reads the compiler's output for the resulting state
and continues from there.

An agent does not only move forward. During proof construction, it may discover
that the current strategy has become unproductive, or that an earlier tactic
quietly led the proof into a dead end. In that case, the agent can backtrack to
the earlier state, revise the proof from that point, and extend a new edge to a
new child. This makes exploration less brittle: the agent does not have to make
a perfect sequence of irreversible choices, but can try a strategy, learn from where
it breaks, and return to a better branching point.

\subsection{Scheduling across Agents}
\label{subsec:scheduling}

Backtracking (\S\ref{subsec:backtracking}) handles the wrong turns that the agent itself recognizes.
It does not address the two limitations observed in Section~\ref{sec:overview}:
an agent may stall at a state instead of backtracking, and a single branch explores alternatives one after another,
even when several strategies are worth exploring together.

To address both limitations, \system{} uses an orchestrator. The orchestrator is
not an LLM agent, but instead a deterministic algorithm that manages the same proof tree from above. It decides where to fork the search with a fresh agent, which branches should be stopped, and how failures on one branch should inform the
others. The orchestrator uses three mechanisms: spawning, pruning, and negative
memory (Fig.~\ref{fig:proof-tree}).

\begin{figure}[t]
  \centering
  \resizebox{\columnwidth}{!}{\begin{tikzpicture}[
  x=1cm,y=1cm,
  font=\sffamily,
  pframe/.style={draw=black!16, fill=softgray, line width=0.5pt,
    rounded corners=2pt},
  state/.style={circle, draw=black!45, fill=white, line width=0.45pt,
    minimum size=2.5mm, inner sep=0},
  oldstate/.style={circle, draw=black!25, fill=white, line width=0.45pt,
    minimum size=2.5mm, inner sep=0},
  stuck/.style={circle, draw=black!45, fill=black!22, line width=0.45pt,
    minimum size=2.5mm, inner sep=0},
  e1/.style={draw=spcyan!65, line width=0.6pt},
  e2/.style={draw=spamber!70, line width=0.6pt},
  e10/.style={draw=spviolet!75, line width=0.7pt},
  eold/.style={draw=black!22, line width=0.55pt},
  edead/.style={draw=spred!60, line width=0.6pt, densely dashed},
  plabel/.style={font=\fontsize{5.6}{6.4}\selectfont\sffamily\bfseries,
    text=black!82, anchor=west},
  ptitle/.style={font=\fontsize{5.2}{6.0}\selectfont\sffamily\bfseries,
    text=black!68, anchor=west},
  note/.style={font=\fontsize{4.4}{5.0}\selectfont\sffamily, text=black!55,
    anchor=west}
]
\definecolor{spgreen}{RGB}{75,135,80}
\definecolor{spamber}{RGB}{187,124,30}
\definecolor{spcyan}{RGB}{55,116,145}
\definecolor{spviolet}{RGB}{128,104,172}
\definecolor{spred}{RGB}{175,72,60}
\definecolor{spgray}{RGB}{122,122,118}
\definecolor{softgray}{RGB}{249,249,247}

\newcommand{\surfaceitem}[2]{%
  \begingroup
  \setlength{\fboxsep}{1.15pt}%
  \setlength{\fboxrule}{0.28pt}%
  \fcolorbox{#1!42}{#1!7}{\textcolor{#1!68!black}{%
    \fontsize{4.75}{5.1}\selectfont\sffamily #2}}%
  \endgroup%
}

\def\PW{3.93}
\def\PH{2.2}

\begin{scope}[shift={(0.12,2.45)}]
  \path[pframe] (0,0) rectangle (\PW,\PH);
  \node[plabel] at (0.14,1.98) {(a)};
  \node[ptitle] at (0.62,1.98) {parallel start};
  \draw[e1] (1.95,1.78) -- (1.30,1.38) -- (0.98,0.98);
  \draw[e2] (1.95,1.78) -- (2.60,1.38) -- (2.92,0.98);
  \foreach \p in {(1.95,1.78),(1.30,1.38),(2.60,1.38),(0.98,0.98),(2.92,0.98)} \node[state] at \p {};
  \node[note] at (2.12,1.80) {initial state};
  \node[inner sep=0] (A1) at (0.98,0.62) {\surfaceitem{spcyan}{Agent 1}};
  \node[inner sep=0] (A2) at (2.92,0.62) {\surfaceitem{spamber}{Agent 2}};
  \draw[e1] (0.98,0.98) -- (A1.north);
  \draw[e2] (2.92,0.98) -- (A2.north);
\end{scope}

\begin{scope}[shift={(4.40,2.45)}]
  \path[pframe] (0,0) rectangle (\PW,\PH);
  \node[plabel] at (0.14,1.98) {(b)};
  \node[ptitle] at (0.62,1.98) {backtrack};
  \draw[e1] (1.95,1.78) -- (1.30,1.38) -- (0.98,0.98);
  \draw[e2] (1.95,1.78) -- (2.60,1.38);
  \draw[eold] (2.60,1.38) -- (2.92,0.98);
  \draw[e2] (2.60,1.38) -- (2.45,0.98);
  \foreach \p in {(1.95,1.78),(1.30,1.38),(2.60,1.38),(0.98,0.98),(2.45,0.98)} \node[state] at \p {};
  \node[oldstate] at (2.92,0.98) {};
  \draw[->, >=stealth, spamber!80, line width=0.5pt, densely dashed]
    (3.02,1.06) to[bend right=50] node[note, text=spamber!85!black, right=1pt] {undo} (2.70,1.42);
  \node[inner sep=0] (A1) at (0.98,0.62) {\surfaceitem{spcyan}{Agent 1}};
  \node[inner sep=0] (A2) at (2.45,0.62) {\surfaceitem{spamber}{Agent 2}};
  \draw[e1] (0.98,0.98) -- (A1.north);
  \draw[e2] (2.45,0.98) -- (A2.north);
\end{scope}

\begin{scope}[shift={(0.12,0.10)}]
  \path[pframe] (0,0) rectangle (\PW,\PH);
  \node[plabel] at (0.14,1.98) {(c)};
  \node[ptitle] at (0.62,1.98) {stall, spawn a child};
  \draw[e1] (1.95,1.78) -- (1.30,1.38) -- (0.98,0.98);
  \draw[e10] (1.30,1.38) -- (1.50,0.98);
  \draw[e2] (1.95,1.78) -- (2.60,1.38) -- (2.45,0.98) -- (2.45,0.58);
  \draw[eold] (2.60,1.38) -- (2.92,0.98);
  \foreach \p in {(1.95,1.78),(1.30,1.38),(2.60,1.38),(1.50,0.98),(2.45,0.98),(2.45,0.58)} \node[state] at \p {};
  \node[oldstate] at (2.92,0.98) {};
  \node[stuck] at (0.98,0.98) {};
  \node[note, text=spgray!90!black, anchor=east, align=right] at (0.84,0.98) {stuck;\\failure\\recorded};
  \node[note, text=spviolet!75!black] at (1.46,1.30) {spawn};
  \node[inner sep=0] (A10) at (1.50,0.62) {\surfaceitem{spviolet}{Agent 1.1}};
  \node[inner sep=0] (A2)  at (2.45,0.24) {\surfaceitem{spamber}{Agent 2}};
  \draw[e10] (1.50,0.98) -- (A10.north);
  \draw[e2]  (2.45,0.58) -- (A2.north);
\end{scope}

\begin{scope}[shift={(4.40,0.10)}]
  \path[pframe] (0,0) rectangle (\PW,\PH);
  \node[plabel] at (0.14,1.98) {(d)};
  \node[ptitle] at (0.62,1.98) {prune stalled branch};
  \draw[e1] (1.95,1.78) -- (1.30,1.38);
  \draw[edead] (1.30,1.38) -- (0.98,0.98);
  \draw[e10] (1.30,1.38) -- (1.50,0.98) -- (1.50,0.58);
  \draw[e2] (1.95,1.78) -- (2.60,1.38) -- (2.45,0.98) -- (2.45,0.58);
  \draw[eold] (2.60,1.38) -- (2.92,0.98);
  \foreach \p in {(1.95,1.78),(1.30,1.38),(2.60,1.38),(1.50,0.98),(1.50,0.58),(2.45,0.98),(2.45,0.58)} \node[state] at \p {};
  \node[oldstate] at (2.92,0.98) {};
  \node[stuck] at (0.98,0.98) {};
  \draw[spred, line width=0.7pt] (0.86,0.86) -- (1.10,1.10);
  \draw[spred, line width=0.7pt] (0.86,1.10) -- (1.10,0.86);
  \node[note, text=spred!80!black, anchor=east] at (0.80,0.98) {pruned};
  \node[inner sep=0] (A10) at (1.50,0.24) {\surfaceitem{spviolet}{Agent 1.1}};
  \node[inner sep=0] (A2)  at (2.45,0.24) {\surfaceitem{spamber}{Agent 2}};
  \draw[e10] (1.50,0.58) -- (A10.north);
  \draw[e2]  (2.45,0.58) -- (A2.north);
\end{scope}

\pgfresetboundingbox
\path[use as bounding box] (0.0,0.0) rectangle (8.45,4.70);
\end{tikzpicture}}
  \caption{A proof tree evolving over a search.
  \textbf{(a)} Two agents explore in parallel from the initial state.
  \textbf{(b)} Agent~2 recognizes a wrong turn, backtracks to an earlier state, and takes a different edge.
  \textbf{(c)} Agent~1 stalls; the orchestrator records the failed attempt and spawns Agent~1.1 at the parent state.
  \textbf{(d)} The stalled branch is pruned ($\times$); Agent~1.1 and Agent~2 continue.}
  \label{fig:proof-tree}
\end{figure}

\smallskip\noindent\textbf{Spawning children.}
At each node, the orchestrator may spawn additional children, each exploring a
different path. Without control, this would quickly become infeasible: spawning
two children from the root, then two children from each child, would make the
number of live agents grow exponentially. \camerareadyonly{\system{} therefore spawns new agents
only when the current search behavior suggests that another strategy is worth
trying.}%
\arxivonly{Most accepted tactics are routine steps, such as simplification, whose alternatives would lead to essentially the same state; branching there only multiplies agents.
The states worth branching from are those where the proof could genuinely go another way: before a tactic that fixes the shape of the rest of the proof, or where an agent is stuck.
The orchestrator therefore creates a new node only at such states.}

\noindent\ding{182}~\emph{Persistent local failure.}
If an agent at a node (state) repeatedly submits tactics and the checker rejects them,
this suggests that the agent may be stuck in a local neighborhood of bad moves and the current branch is unproductive.
Rather than letting the agent spend more time on the same strategy, the
orchestrator creates a sibling branch. It does so by spawning a child {\em at the
parent} of the node where the agent is currently working, giving the search a
chance to approach the state through a different path.

\noindent\ding{183}~\emph{Structural undo.}
Some tactics fix the shape of the rest of the proof: they open a procedure or turn a probability statement into a program equivalence (e.g., \texttt{proc}, \texttt{byequiv}),
cut the program with an intermediate assertion or invariant (\texttt{seq}, \texttt{while}, \texttt{call}), inline code (\texttt{inline}),
or introduce an intermediate fact (\texttt{have}, \texttt{transitivity}).
We call an undo \emph{structural} when it undoes such a tactic:
the agent has abandoned an earlier decision that shaped much of the later proof, so the state before that tactic is a valuable branching point.
If the same agent has not repaired its strategy within five minutes, the orchestrator spawns a child at that state.

\smallskip\noindent\textbf{Pruning.}
Spawning creates new opportunities, but it also creates the risk of an
unbounded search. The orchestrator therefore prunes branches and terminates
unproductive agents. The pruning decision is based on a \emph{progress gap}: if
an agent remains far behind the other agents in the tree, it becomes a candidate
for pruning.

This decision has to be made carefully. A branch that looks behind is not necessarily useless: an agent may be backtracking through earlier states, which temporarily reduces its depth but may eventually lead to a better proof strategy. The orchestrator therefore distinguishes agents that are temporarily behind because they are performing meaningful backtracking from agents that
remain behind throughout the search. In this way, pruning keeps the number of
live agents bounded without discarding every branch that briefly moves backward.

\smallskip\noindent\textbf{Negative memory.}
When one agent fails to extend a path, the other agents should not blindly
repeat the same experiment. The orchestrator therefore records failures as
state-indexed \emph{negative memory}. For a given node, negative memory records
which attempted extension led to a dead end and why. When the orchestrator later
spawns a child at that node, it passes the corresponding negative memory to the
new agent, so that the new branch can avoid spending budget on the same failed
strategy.

Negative memory is a search-control mechanism, not proof knowledge. It records
that a particular experiment failed at a particular state; it does not
claim that the underlying proof strategy is impossible.

\section{Coordinating Proof Decomposition and Lemma Proving}
\label{sec:coordination}

In Sections~\ref{sec:proof-state-compiler} and~\ref{sec:tree-policy}, the lemma prover proves lemmas whose statements are already given.
We now start from only the specification and the final theorem, whose proof is left open,
and let \system{} construct the decomposition as well (Phase~II in Fig.~\ref{fig:running-example}).
Phase~II need not finish before Phase~III starts: whether a decomposition is good is often known only after trying to prove its lemmas.
For example, an invariant may turn out to be too weak to carry through an oracle call, or a planned game hop may violate the restrictions on which modules the adversary may access.
As experts do, \system{} moves back and forth between the two phases (Fig.~\ref{fig:coordination}).

\smallskip\noindent\textbf{Decomposer.}
The decomposer is an LLM agent that works on an EasyCrypt file containing the specification and the statement of the final theorem.
Its goal is to complete the proof of the final theorem in this file, so that EasyCrypt accepts the whole file with no \texttt{admit} remaining.
In our experiments, the final theorem and the supplied definitions must remain unchanged,
and new declarations may be added only in a marked region of the file,
so the decomposer cannot make the theorem easier by weakening it or adding assumptions.
The decomposer chooses the intermediate games, writes the auxiliary lemmas together with the modules and invariants they need,
and assembles the proof of the final theorem; it may also prove lemmas directly.
It checks its work by replaying the whole file in EasyCrypt, temporarily allowing \texttt{admit} placeholders,
so that it can confirm that a skeleton of the argument type-checks and connects to the final theorem before every lemma is proved.

\smallskip\noindent\textbf{Handing lemmas to the lemma prover.}
When a lemma would benefit from step-by-step interaction with the proof state, the decomposer hands it to the lemma prover as a \emph{job}.
A job works on exactly one lemma, using the proof-state compiler and the proof tree of \S\ref{sec:proof-state-compiler} and \S\ref{sec:tree-policy}.
The decomposer never hands over the final theorem or asks the lemma prover to invent the overall reduction:
choosing game hops and lemma statements remains its own responsibility.
Each job receives a snapshot of the file, the name of the lemma, a time limit, and usually a short note
with the goal, any proof prefix already accepted by EasyCrypt, the intended strategy, and the attempts that have already failed.
A job that receives an accepted prefix first replays it, so it never starts from a state that the checker has not accepted.
Depending on the progress already accepted by EasyCrypt, the decomposer requests one of three kinds of job:
an exploratory job that tests whether a game hop, an invariant, or a proof strategy works;
a completion job once accepted tactics already fix the strategy;
or a continuation of an earlier job from its saved state.
Handing off a lemma does not block the decomposer: a limited number of jobs run at the same time,
while the decomposer continues with independent parts of the argument, such as later games or probability bounds.

\smallskip\noindent\textbf{Feedback from Phase~III to Phase~II.}
A job that proves its lemma returns the proof, which is checked again against the current file before it is merged.
Otherwise, the job returns the longest prefix of tactics that EasyCrypt accepted, the remaining goals,
and, when available, a saved state for a continuation; any notes it adds are hints, not checked facts.
From such a result, the decomposer can finish the proof itself, request a continuation,
restate or split the lemma or add helper lemmas, or change the decomposition.
The last two are where Phase~III feeds back into Phase~II (Fig.~\ref{fig:coordination}; \S\ref{subsec:eval-case} describes an example from the ChaChaPoly case study).

\section{Evaluation}
\label{sec:evaluation}

\smallskip\noindent\textbf{Implementation.}
\system{} is an open-source research prototype written in Python and built on EasyCrypt. The
core implementation contains about 89K lines of code: about 86K lines of Python and about 3K lines of OCaml that extend EasyCrypt.
By component, it comprises about 33K lines for the multi-agent tree search (\S\ref{sec:tree-policy}), including the orchestrator, the agents,
and recovery from agent crashes; about 26K lines for the proof-state compiler (\S\ref{sec:proof-state-compiler});
about 24K lines for running EasyCrypt and parsing its output; and about 7K lines for the decomposer (\S\ref{sec:coordination}).
App.~\ref{app:system-architecture} describes the architecture in detail.
Our code is available at \url{https://github.com/SkyShannonProver/shannon-prover}.

\subsection{Evaluation Design}
\label{subsec:eval-setup}

Our evaluation has two parts.
The first is a controlled ablation of the lemma prover's two design ideas.
For the proof-state compiler, we measure how much it reduces the agent's errors and the cost of recovering from them.
For the multi-agent tree search, we measure how much it speeds up proof writing compared with a single agent.
The second part is a case study on ChaCha20-Poly1305~\cite{almeida2020lastmile},
in which we evaluate both the lemma prover given the experts' decomposition and the end-to-end setting with the decomposer.
Appendix~\ref{app:crypto-benchmark} presents additional case studies on
MEE-CBC~\cite{almeida2016meecbc} and CMAC~\cite{baritel2018formal},
together with expert comments on these developments (App.~\ref{app:expert-comments}).

\smallskip\noindent\textbf{Setup.}
All ablations share the following setup.
We use Opus~4.8\footnote{This was the strongest model available when we ran the ablation experiments.} with high reasoning effort.
The agent may read all project files and \easycrypt{} source files, and we give it no hints about proof strategy.
Each ablation compares the lemma prover with a baseline in which the mechanism under study is disabled
(\S\ref{subsec:eval-compiler} and \S\ref{subsec:eval-tree} describe each baseline).
The baseline uses the same MCP-based \easycrypt{} interaction layer as the lemma prover,
so that differences in results do not stem from unrelated engineering factors such as session management and crash recovery.

We use the same stopping rule in all configurations: the agent may announce that it gives up at most three times. After each announcement, the system prompts the agent to try another strategy. If the agent makes three consecutive such announcements without proof progress, we count the run as a give-up. This rule prevents runs from spending an unbounded budget in states where the agent repeatedly reports that it is stuck. In pilot runs, forcing the agent to continue led either to wasted tokens at the same stuck point or to behavior unrelated to the proof, such as reading files outside the project.

\smallskip\noindent\textbf{Data sampling.}
To keep the API cost of the ablation study manageable while keeping the evaluation representative,
we sample 40 lemmas from our benchmark dataset, covering different types and difficulty levels. Appendix~\ref{app:crypto-benchmark} describes how we built this dataset.
We group lemmas by proof shape: procedural (\textbf{P}), invariant-synthesis (\textbf{I}), and game-hop/reduction (\textbf{G}). This grouping is not meant to label cryptographic importance; rather, it separates qualitatively different proof-engineering regimes. Procedural lemmas are mostly {\em local}: once the relevant procedure body is exposed, progress is often made by a sequence of greedily discoverable symbolic-execution and simplification tactics. Invariant-synthesis lemmas require the agent to choose a relational invariant that is strong enough to carry through a call or loop boundary, but not so strong that it creates unnecessary side obligations. Game-hop and reduction lemmas require the proof to preserve higher-level game structure long enough to apply bridge lemmas, oracle equivalences, or probability bounds. These three shapes pose different difficulties for the agent, so aggregating them into a single average would obscure where the lemma prover helps.

\smallskip\noindent\textbf{Metrics.}
LLM-based theorem provers are usually evaluated by pass@$k$~\cite{polu2020gptf,zheng2021minif2f}:
the model writes $k$ complete proof scripts, and a theorem counts as solved if any of them is accepted.
We do not use this metric, for two reasons.
First, it suits short proofs that can be written in one shot, as in miniF2F~\cite{zheng2021minif2f}.
Our lemmas often need dozens of tactics, so writing a complete proof in one shot that the checker accepts is unrealistic,
and sampling many attempts (i.e., a very large $k$) is prohibitively expensive.
Second, the binary outcome, whether a lemma is proved or not, does not show what our mechanisms change.
For example, two runs may both succeed, one through a short, direct proof and the other only after many rejected tactics.
We instead use metrics that target each mechanism, detailed in \S\ref{subsec:eval-compiler} and \S\ref{subsec:eval-tree}.

\subsection{The Effects of the Proof-State Compiler}
\label{subsec:eval-compiler}

\smallskip\noindent\textbf{Baseline.}
The baseline has the agent interact directly with the checker: after submitting each tactic to the checker
it sees only the resulting goal or the error message. The lemma prover
instead shows the agent the output of the proof-state compiler
(\S\ref{sec:proof-state-compiler}).

\smallskip\noindent\textbf{Metrics.}
A cheaper proof could come from two very different places, and a fair ablation
has to separate them. The compiler's output might make the model propose fewer
wrong tactics in the first place, or it might leave the first-attempt error
rate unchanged while making those errors far less costly. We therefore have the following metrics:

\noindent\ding{182}~\emph{Error-generation rate.}
The fraction of the model's distinct proposed tactics that the checker rejects
on its first attempt.

\noindent\ding{183}~\emph{Error friction.}
The fraction of an agent's total API spend on sequences of consecutive tactics
rejected by the checker.
This captures the fraction that the agent spends on debugging syntax and recovering from errors.

To collect these metrics, our measurement pipeline separates deterministic metrics from qualitative judgement.
The first stage is a Python parser:
for each run it extracts the submitted tactics,
the checker responses,
and the agent reasoning traces for each submitted tactic.
The second stage uses an LLM for cases a parser cannot classify mechanically: for example, when the agent
commits several steps and later reverts them with no explicit checker error, so
the parser cannot tell an unproductive dead end from a deliberate
restructuring; there the LLM labels those tactic moves with the corresponding qualitative cause. The final stage
is deterministic again: a script aggregates the parsed events and the
labels into the reported numbers.

\begin{figure}[t]
  \centering
  \begin{tikzpicture}
\begin{groupplot}[
  group style={group size=2 by 3, horizontal sep=20pt, vertical sep=14pt},
  ablpanel,
  width=0.57\columnwidth,
  xmin=-0.62, xmax=1.62,
  xtick={0,1},
  xticklabels={,},
  ybar, /pgf/bar width=17pt,
  every axis plot/.append style={/pgf/bar shift=0pt},
]
\nextgroupplot[title={(a) Error-generation\\rate}, ylabel={\textbf{P}}, ylabel style={rotate=-90}, ymax=0.137]
\addplot[ybar, fill=ablbase, draw=none] coordinates {(0,0.100)};
\addplot[ybar, fill=ablours, draw=none] coordinates {(1,0.068)};
\node[ablbarlabel, above] at (axis cs:0,0.100) {0.10};
\node[ablbarlabel, above] at (axis cs:1,0.068) {0.07};
\nextgroupplot[title={(b) Error friction\\(\% of spend)}, ymax=14.8]
\addplot[ybar, fill=ablbase, draw=none] coordinates {(0,10.8)};
\addplot[ybar, fill=ablours, draw=none] coordinates {(1,0.02)};
\node[ablbarlabel, above] at (axis cs:0,10.8) {10.8\%};
\node[ablbarlabel, above] at (axis cs:1,0.02) {0.0\%};
\nextgroupplot[ylabel={\textbf{I}}, ylabel style={rotate=-90}, ymax=0.166]
\addplot[ybar, fill=ablbase, draw=none] coordinates {(0,0.083)};
\addplot[ybar, fill=ablours, draw=none] coordinates {(1,0.121)};
\node[ablbarlabel, above] at (axis cs:0,0.083) {0.08};
\node[ablbarlabel, above] at (axis cs:1,0.121) {0.12};
\nextgroupplot[ymax=16.9]
\addplot[ybar, fill=ablbase, draw=none] coordinates {(0,12.3)};
\addplot[ybar, fill=ablours, draw=none] coordinates {(1,0.3)};
\node[ablbarlabel, above] at (axis cs:0,12.3) {12.3\%};
\node[ablbarlabel, above] at (axis cs:1,0.3) {0.3\%};
\nextgroupplot[ylabel={\textbf{G}}, ylabel style={rotate=-90}, ymax=0.184,
  xticklabels={{baseline},{\system{}}}]
\addplot[ybar, fill=ablbase, draw=none] coordinates {(0,0.110)};
\addplot[ybar, fill=ablours, draw=none] coordinates {(1,0.134)};
\node[ablbarlabel, above] at (axis cs:0,0.110) {0.11};
\node[ablbarlabel, above] at (axis cs:1,0.134) {0.13};
\nextgroupplot[ymax=13.4, xticklabels={{baseline},{\system{}}}]
\addplot[ybar, fill=ablbase, draw=none] coordinates {(0,9.8)};
\addplot[ybar, fill=ablours, draw=none] coordinates {(1,1.3)};
\node[ablbarlabel, above] at (axis cs:0,9.8) {9.8\%};
\node[ablbarlabel, above] at (axis cs:1,1.3) {1.3\%};
\end{groupplot}
\end{tikzpicture}
  \caption{Compiler ablation on lemma pairs that both cases proved, grouped by
  proof shape: \textbf{P}~= procedural, \textbf{I}~= invariant-synthesis, \textbf{G}~=
  game-hop/reduction.}
  \label{fig:compiler-ablation}
\end{figure}

\smallskip
\noindent
\textbf{Results.}
Figure~\ref{fig:compiler-ablation} shows that the proof-state compiler does not primarily work by making the model intrinsically less error-prone. The error-generation rate is close between the baseline and the lemma prover across the three lemma types.
This is expected, because this metric is dominated by the model's own stochastic search behavior.

The error-friction numbers show the more important effect. Across all lemma types, the lemma prover sharply reduces the fraction of spend on such rejected sequences. This means that even when the agent still makes mistakes when it first tries a move, the agent can immediately recover from it.
Meanwhile, the agent in the baseline must figure out the right move via trial-and-error. The ablation shows that the proof-state compiler is not a proof oracle that provides the right move, but an interface that provides sufficient information to free the agent from mechanical debugging.

The main counter-intuitive result for the error-generation rate is invariant-synthesis lemmas, where the lemma prover has a visibly higher rate.
The experimental traces suggest why: with the richer information provided by the compiler, the agent often attempts stronger and more structured invariants, because it can see more of the resources that apply, the program statements to which they apply, and the required bindings.
Some of these ambitious invariant attempts are rejected on the first attempt.
In the baseline, however, the agent is more likely to choose a weaker invariant that passes immediately, even if it leaves the proof in a worse state later.

The compiler ablation therefore shows that the proof-state compiler mainly reduces the cost of errors rather than their number:
the agent still makes mistakes, but it spends far less of its budget recovering from them.

\subsection{The Effects of Multi-Agent Tree Search}
\label{subsec:eval-tree}

\smallskip\noindent\textbf{Baseline.}
The multi-agent tree search has two layers (\S\ref{sec:tree-policy}): backtracking within a branch, and spawning further branches from shared checkpoints.
Backtracking itself cannot be meaningfully removed: an agent that cannot undo a wrong step can only fail.
The baseline therefore keeps backtracking and removes spawning.
It runs one branch at a time: the agent can undo and backtrack within its branch, and if its agent process crashes, the branch is resumed from its committed prefix, but no sibling branch ever races it.
The lemma prover adds the cross-agent tree policy: a second branch can race from a shared checkpoint, and dead or stalled nodes can be respawned with negative memory so that the next agent does not repeat the same failed strategy.
In both configurations, the agent can use the proof-state compiler.

\smallskip\noindent\textbf{Metrics.}
We measure how often the agent needs to backtrack: the fraction of solved runs that backtrack at least once, and the number of backtrack events per solved run.
These numbers show how far proof search departs from a straight line in each lemma category.
We also measure the solve rate in each category, and how it grows over the course of a run.

\smallskip\noindent\textbf{Results.}
Figure~\ref{fig:tree-ablation}a--b first shows that the need for backtracking is highly lemma-type dependent. Procedural lemmas are almost straight-line, as expected from their local structure.
In contrast, invariant-synthesis and game-hop lemmas are much less linear.
38\% of solved invariant runs and 43\% of solved game-hop runs backtrack at least once, with about 2.0--2.3 backtrack events per solved run. These are exactly the lemmas where an attempted tactic that is accepted by the checker can lead the agent into an unproductive strategy: an invariant may be too weak or too strong, and a game-hop proof may descend too early and lose a bridge lemma or oracle-equivalence resource. For these lemmas, the ability to rewind is not a rare recovery feature; it is part of how successful proofs are found.

Figure~\ref{fig:tree-ablation}c--d shows what spawning adds on top of an agent that backtracks.
On procedural lemmas, backtracking within one branch already suffices: the single agent proves every lemma, and a second branch only wins the race.
On invariant-synthesis and game-hop lemmas, it does not: even with backtracking, a third of single-agent runs fail,
and spawning raises the solve rate from 67\% to 90\%.
The gain comes mainly from redundancy: the second branch completed the proof after the root branch's agent had died.
It also appears late (Fig.~\ref{fig:tree-anytime}): both configurations solve 82\% of all runs within 15 minutes and separate only afterwards, once a first branch has stalled.

The tree ablation therefore supports two conclusions. First, proof search is not uniformly non-linear: for local procedural obligations, a single branch with backtracking is enough.
Second, for the proof types that matter most in cryptographic developments---invariants and game hops---backtracking within one branch is not enough, and spawning further branches makes the search robust to branches that stall or die.

\begin{figure}[t]
  \centering
  \begin{tikzpicture}
\begin{groupplot}[
  group style={group size=2 by 2, horizontal sep=30pt, vertical sep=52pt},
  ablpanel,
  width=0.565\columnwidth,
  ybar, /pgf/bar width=14pt,
  every axis plot/.append style={/pgf/bar shift=0pt},
]
\nextgroupplot[title={(a) Solved runs that\\backtracked},
  ylabel={\% of solved runs}, ymax=52,
  xmin=-0.62, xmax=2.62, xtick={0,1,2},
  xticklabels={{\textbf{P}},{\textbf{I}},{\textbf{G}}}]
\addplot[ybar, fill=ablgrnA, draw=none] coordinates {(0,3)};
\addplot[ybar, fill=ablgrnB, draw=none] coordinates {(1,38)};
\addplot[ybar, fill=ablgrnC, draw=none] coordinates {(2,43)};
\node[ablbarlabel, above] at (axis cs:0,3) {3\%};
\node[ablbarlabel, above] at (axis cs:1,38) {38\%};
\node[ablbarlabel, above] at (axis cs:2,43) {43\%};
\nextgroupplot[title={(b) Backtrack events\\per solved run},
  ylabel={events\,/\,solved run}, ymax=2.65,
  xmin=-0.62, xmax=2.62, xtick={0,1,2},
  xticklabels={{\textbf{P}},{\textbf{I}},{\textbf{G}}}]
\addplot[ybar, fill=ablgrnA, draw=none] coordinates {(0,0.04)};
\addplot[ybar, fill=ablgrnB, draw=none] coordinates {(1,2.00)};
\addplot[ybar, fill=ablgrnC, draw=none] coordinates {(2,2.29)};
\node[ablbarlabel, above] at (axis cs:0,0.04) {0.04};
\node[ablbarlabel, above] at (axis cs:1,2.00) {2.00};
\node[ablbarlabel, above] at (axis cs:2,2.29) {2.29};
\nextgroupplot[title={(c) Winning path needed\\backtrack\,/\,respawn},
  ylabel={\% of orch.\ solves}, ymax=94,
  xmin=-0.62, xmax=2.62, xtick={0,1,2},
  xticklabels={{\textbf{P}},{\textbf{I}},{\textbf{G}}}]
\addplot[ybar, fill=ablgrnA, draw=none] coordinates {(0,3)};
\addplot[ybar, fill=ablgrnB, draw=none] coordinates {(1,80)};
\addplot[ybar, fill=ablgrnC, draw=none] coordinates {(2,50)};
\node[ablbarlabel, above] at (axis cs:0,3) {3\%};
\node[ablbarlabel, above] at (axis cs:1,80) {80\%};
\node[ablbarlabel, above] at (axis cs:2,50) {50\%};
\nextgroupplot[title={(d) Solve rate\\(\textbf{I} $+$ \textbf{G} lemmas)},
  ylabel={\% of runs solved}, ymax=112,
  xmin=-0.62, xmax=1.62, xtick={0,1}, /pgf/bar width=17pt,
  xticklabels={{single\\agent},{orches-\\trator}}]
\addplot[ybar, fill=ablbase, draw=none] coordinates {(0,67)};
\addplot[ybar, fill=ablours, draw=none] coordinates {(1,90)};
\node[ablbarlabel, above] at (axis cs:0,67) {67\%};
\node[ablbarlabel, above] at (axis cs:1,90) {90\%};
\end{groupplot}
\end{tikzpicture}
  \vspace{-18pt}
  \caption{Tree-search ablation, grouped by proof shape
  (\textbf{P}~= procedural;
  \textbf{I}~= invariant-synthesis; \textbf{G}~=
  game-hop/reduction).}
  \label{fig:tree-ablation}
\end{figure}

\begin{figure}[t]
  \centering
  \begin{tikzpicture}
\begin{axis}[
  ablpanel,
  width=0.95\columnwidth, height=3.0cm,
  xmin=0, xmax=70, ymin=60, ymax=106,
  xtick={0,15,30,45,60}, ytick={60,80,100},
  xlabel={elapsed time (minutes)}, ylabel={\% of runs solved},
  clip=true,
]
\addplot[const plot, line width=0.8pt, draw=ablbase] coordinates {(0.00,0.0) (0.12,2.3) (0.12,4.5) (0.14,6.8) (0.14,9.1) (0.15,11.4) (0.15,13.6) (0.16,15.9) (0.19,18.2) (0.20,20.4) (0.40,22.7) (0.69,25.0) (0.72,27.3) (0.75,29.5) (0.83,31.8) (0.88,34.1) (0.90,36.4) (1.22,38.6) (1.37,40.9) (1.40,43.2) (1.64,45.5) (2.00,47.7) (2.10,50.0) (2.11,52.3) (2.16,54.5) (2.25,56.8) (2.46,59.1) (2.58,61.4) (2.74,63.6) (2.76,65.9) (2.97,68.2) (3.62,70.5) (3.80,72.7) (6.52,75.0) (9.21,77.3) (11.13,79.5) (11.39,81.8) (42.16,84.1) (44.21,86.4) (67.99,88.6) (70.00,88.6)};
\addplot[const plot, line width=0.8pt, draw=ablours] coordinates {(0.00,0.0) (0.12,2.3) (0.12,4.5) (0.12,6.8) (0.12,9.1) (0.16,11.4) (0.16,13.6) (0.17,15.9) (0.18,18.2) (0.21,20.4) (0.58,22.7) (0.65,25.0) (0.77,27.3) (0.95,29.5) (1.15,31.8) (1.26,34.1) (1.30,36.4) (1.35,38.6) (1.55,40.9) (1.72,43.2) (1.76,45.5) (1.86,47.7) (1.88,50.0) (1.88,52.3) (1.88,54.5) (2.08,56.8) (2.16,59.1) (2.26,61.4) (2.33,63.6) (2.41,65.9) (2.43,68.2) (2.77,70.5) (3.06,72.7) (3.37,75.0) (4.98,77.3) (5.10,79.5) (7.41,81.8) (19.43,84.1) (32.85,86.4) (43.75,88.6) (44.85,90.9) (46.59,93.2) (63.19,95.5) (68.50,97.7) (70.00,97.7)};
\draw[black!30, densely dashed, line width=0.4pt] (axis cs:15,60) -- (axis cs:15,106);
\node[font=\scriptsize, text=ablours!80!black, anchor=south] at (axis cs:55,93.6) {orchestrator};
\node[font=\scriptsize, text=ablbase!80!black, anchor=north] at (axis cs:55,85.8) {single agent};
\end{axis}
\end{tikzpicture}
  \vspace{-8pt}
  \caption{Fraction of lemmas solved by elapsed time.
  The two configurations coincide for the first 15 minutes (dashed line) and separate afterwards.}
  \label{fig:tree-anytime}
\end{figure}

\subsection{Case Studies}
\label{subsec:eval-case}

ChaCha20-Poly1305~\cite{almeida2020lastmile} is a representative case study project.
It is an AEAD construction that combines the ChaCha20 stream cipher with the Poly1305 authenticator and is used in IETF protocols and deployed in TLS~\cite{rfc8446}. Its EasyCrypt formal proof contains a mixture of low-level algebraic reasoning, invariant-heavy relational proofs, and game-hop/reduction lemmas.
Experts described the development as interleaving the design of new abstractions with mechanical proof engineering: abstraction choices shaped the proof structure, while mechanization difficulties fed back into the abstraction design.

\begin{figure*}[t]
  \centering
  \includegraphics[width=\textwidth]{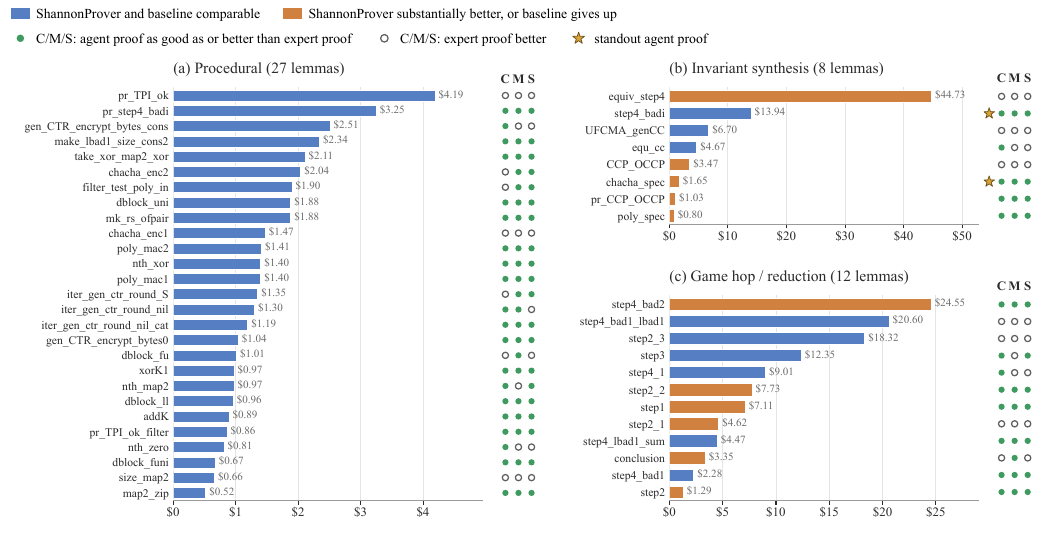}
  \caption{Per-lemma API cost for the ChaCha20-Poly1305 project, grouped by proof shape (\textbf{P}/\textbf{I}/\textbf{G}). Bar color compares \system{} with the baseline. For concision (\textbf{C}), maintainability/readability (\textbf{M}), and semantic proof idiom (\textbf{S}), a filled green dot denotes an agent proof comparable to or better than the expert proof, while a hollow gray circle denotes an expert advantage. Stars mark the two agent proofs described in Appendix~\ref{app:failures-and-standouts}.}
  \label{fig:chachapoly-per-lemma}
\end{figure*}

This case study demonstrates that LLM agents today can fully automate lemma-level proof (Phase III in Fig.~\ref{fig:running-example}) at the scale  and complexity of ChaChaPoly.
Figure~\ref{fig:chachapoly-per-lemma} shows all the lemmas in ChaChaPoly and the API cost per lemma.
The procedural lemmas are generally easy and they are solved at low API cost. This is already useful because large projects contain many such lemmas, and writing them manually still consumes expert time. Importantly, \system{} solved the hardest lemmas in the project, while the baseline agent failed.
These hard lemmas are fewer in the project, but they are also the most important: they are the points where the lemma decomposition is tested. If these lemmas can be proven, the expert gains evidence that the decomposition has the right structure. If LLM agents or humans fail to prove a lemma, the thought process in the failed attempts is still informative: it often indicates that the current lemma is missing an invariant, that a game hop has not been modularized at the right boundary, or that some other part of the decomposition needs to be revised. This feedback is valuable for redoing the decomposition.

We also evaluate proof quality on three fronts:
 concision (\textbf{C}),  maintainability/readability (\textbf{M}) and semantic proof idiom (\textbf{S}). 
 \textbf{C} measures whether the proof
is compact and avoids unnecessary tactic churn or duplicated reasoning.
\textbf{M} measures whether the proof structure
is easy for a human to inspect, debug, and adapt when
the surrounding definitions change. 
\textbf{S} measures whether the
proof follows the natural EasyCrypt proof structure and uses
existing proven facts, rather than reproving low-level details.
Two experts independently rated each pair of proofs, presented in random order, on each dimension. 
Figure~\ref{fig:chachapoly-per-lemma} shows that the AI-generated proofs still leave room for improvement on invariant and game-hop lemmas, making proof quality a promising direction for future work.

The API cost is also low enough to make iterative proof development practical. Across the ChaCha20-Poly1305 lemmas (Fig.~\ref{fig:chachapoly-per-lemma}), proving all lemmas of the expert decomposition once costs \$233.65 for API calls and completes within a day of wall-clock time. The cost is concentrated in the harder lemmas: the 27 procedural lemmas together cost \$40.98, while the invariant-synthesis and game-hop/reduction lemmas account for \$76.99 and \$115.68, respectively. This distribution is desirable from a workflow perspective: the system spends little on procedural lemmas, and most of the cost goes to the lemmas that are most informative for testing the decomposition. Even if a cryptographer revises the decomposition and reruns the proof pass several times, the marginal cost remains on the order of a few hundred dollars per iteration and the time is at most a day, rather than person-weeks or person-months of expert proof engineering. This makes ShannonProver useful not only as a one-shot proof generator, but also as a practical tool for iterating on large cryptographic formalizations.

\smallskip
\noindent
\textbf{Evaluation for end-to-end proof.}
We further evaluate \system{} in an end-to-end setting, where the system is given only the specification 
and the statement of the main theorem. 
In this workflow, the complete proof was constructed in 5 hours, with Opus~5\footnote{We ran this experiment during the revision of this paper, after the other ablations; Opus~5 was the strongest model available at that time.} as the decomposer and the same lemma prover as before.
In this run, about 20\% of the lemmas in the decomposition were reworked after the decomposer first wrote them and
received feedback from the lemma prover.
The total cost is about \$420 for a complete proof.  
In comparison, a baseline that runs Opus 5 through its CLI takes 8 hours to complete the proof end to end, and Opus 4.8 failed to complete it.

\section{Related Work}

\smallskip\noindent\textbf{Machine-checked cryptography.}
Tools for machine-checked cryptography trade expressiveness for automation.
Symbolic tools such as Tamarin~\cite{meier2013tamarin} and ProVerif~\cite{blanchet2016proverif}
are the most automatic: they model the adversary as a Dolev-Yao attacker~\cite{dolev1983security}
and treat cryptographic primitives as perfect, so they cannot state how likely an attack succeeds.
CryptoVerif~\cite{blanchet2008cryptoverif} brings automation to the computational model,
but it builds proofs from a fixed set of game transformations and library assumptions;
arguments outside this set, such as a new reduction or reasoning about the mathematics of a scheme,
need extensive manual guidance or cannot be carried out.
Proof assistants such as EasyCrypt~\cite{barthe2011workingcryptographer} and others~\cite{barthe2009certicrypt, basin2019crypthol, abate2021ssprove, petcher2015fcf, tuma2026vcvio}
can express any reduction together with the mathematics it relies on,
and experts have used them to verify standardized schemes and NIST post-quantum standards~\cite{almeida2024mlkem, barbosa2024sphincs, barbosa2023xmss, almeida2020lastmile, almeida2016meecbc, almeida2019sha3},
in some cases down to verified implementations~\cite{zinzindohoue2017hacl, protzenko2020evercrypt, almeida2017jasmin};
in exchange, the user writes every proof step.
LLM agents change this trade-off: they can carry out the many detailed steps that a proof assistant leaves to the user,
so automation no longer requires giving up expressiveness.

Among proof assistants, we build on EasyCrypt: it supports the full range of reasoning in game-based proofs,
and the expert developments in our case studies were originally written in it.
Lean has a much larger ecosystem and is the target of most LLM-based theorem proving,
but game-based proofs there first need a library for probabilistic programs with oracles.
VCVio~\cite{tuma2026vcvio} provides such a library, with oracle computations and a relational program logic;
it is a promising foundation, but it is still under active development, and its core definitions and semantics change frequently.
We plan to port \system{} to Lean as such libraries mature.

\smallskip\noindent\textbf{LLMs for formal proofs.}
LLM-based theorem proving has advanced rapidly, mostly in mathematics,
through whole-proof generation~\cite{polu2020gptf, lample2022htps, xin2024deepseekproverv15, ren2025deepseekproverv2, lin2025goedelprover, deepmind2025alphaproof}
evaluated on competition benchmarks~\cite{zheng2021minif2f, azerbayev2023proofnet, tsoukalas2024putnambench},
tactic-level interaction with the proof assistant~\cite{yang2023leandojo, jiang2022thor, first2023baldur},
and, more recently, agentic provers~\cite{breen2025axprover, sosso2026agentic}.
In cryptography, reinforcement learning and learned search heuristics have guided symbolic protocol proofs~\cite{xiong2020smartverif, cosler2026tamarinrl},
LLMs have been used to build or check protocol models~\cite{curaba2024cryptoformaleval, mao2025llmaided},
generative AI has assisted a researcher in formalizing computational security proofs in Lean~\cite{spitters2026catcrypt},
and AI provers have discharged correctness obligations extracted from cryptographic Rust code~\cite{klaus2026rustlean}.
Other work enriches the interface between model and prover,
by suggesting tactics in the human workflow~\cite{song2024leancopilot, welleck2023llmstep}, exposing proof states to agents~\cite{aniva2024pantograph},
or adding a natural-language representation of the proof state~\cite{teodorescu2024nlir}.

\section{Conclusion and Future Work}

This work shows that LLM agents can take on a substantial part of mechanized cryptographic proof
when the system around them supplies the right context and controls the search.

Two problems remain open.
Decomposing a large proof obligation into provable lemmas (Phase II) still depends strongly on the capability of the model;
better harness support for the decomposer may improve performance.
For Phase I, AI could help experts not only write the specification, but also define the semantic abstractions a proof needs.
Such abstractions are often discovered only while writing the proof, when a lemma turns out to be hard to state or prove with the existing definitions;
supporting them requires moving back and forth between Phase I and the later phases, as experts do.

\section{Generative AI Responsibility Policy}
\label{sec:generative-ai-responsibility}

\noindent 
\textbf{Content and data holder rights.}
Our benchmark and case studies are built from EasyCrypt developments (App.~\ref{app:crypto-benchmark}), and we used each in a manner consistent with the terms under which its rights holders released it.

The foundational primitives, MEE-CBC, and ChaCha20-Poly1305 come from the example suite distributed with EasyCrypt under the MIT License. The ML-KEM development is released under the Creative Commons CC0 1.0 Universal public domain dedication~\cite{almeida2024mlkemArtifact}, the SLH-DSA/SPHINCS$^{+}$ development under the MIT License, and the XMSS development under the Apache License 2.0. The ML-DSA/Dilithium development is publicly available without an explicit license; we use it only for evaluation and do not redistribute it. 

The CMAC development is not publicly available. We obtained explicit authorization from its authors to evaluate ShannonProver on it, and our use is confined to the scope of that authorization. We therefore report aggregate results for CMAC without releasing the underlying source files; readers seeking access should contact the original authors directly.

\smallskip 
\noindent 
\textbf{Environmental footprint.}
ShannonProver substantially reduces token usage compared with using Codex or Claude through a plain CLI to solve the same task. We believe this points to a more responsible way of using LLMs for proof automation going forward: whenever possible, work that does not require substantial model reasoning should be moved to deterministic components, reserving LLM calls for tasks where their reasoning capabilities are genuinely needed.

\section*{Acknowledgments}

This work is supported by gifts from Accenture, Algorithmic SuperIntelligence Labs, Amazon, AMD, Anyscale, Broadcom, cmpnd, Google, IBM, Intel, Intesa Sanpaolo, Lambda, Lightspeed, Mirendil, NVIDIA, Samsung SDS, and VESSL.

We would like to thank Quang Dao, Shafi Goldwasser, Mohsen Lesani, Hongyi Ling, Ning Luo,
Surya Mathialagan, Yibin Yang, the students in the security group at Sky Lab, and the participants and organizers in HACS 2026 for valuable discussions in the early stages of this project. We also thank Sebastian Angel, Shai Halevi, Yuval Ishai, and Tal Rabin for feedback and discussion on future directions.

\section*{Competing Interests}

The research funding listed in the acknowledgments consists of unrestricted gifts, and the funders had no role in the design, analysis, or writing of this work. The authors declare no other competing financial or non-financial interests.


\appendices
\arxivonly{\section{Deferred Explanations of Concepts}
\label{app:deferred-concept-explanations}

\subsection{Proof Resources}
\label{app:proof-resources}

A \emph{proof resource} is something the proof can use instead of reasoning
from scratch. The main kinds are lemmas already proved in the project or in a
library (including equivalence lemmas, bridge lemmas, and probability bounds),
axioms and assumptions of the specification, and hypotheses in the current
proof. Each of them packages reasoning that has already been done or assumed;
using it at the right point saves the proof from redoing that reasoning.

Two properties of resources matter for \system{}. First, resources are chosen,
not invented: the agent decides which resource to use, whereas an invariant or
a new lemma has to be created before it can be used. Second, which resources
can be used changes as the proof proceeds. Choosing a resource is a strategic
decision left to the agent; finding the resources that apply at the current
state and instantiating them is mechanical work done by the compiler
(\S\ref{sec:proof-state-compiler}). Filling a premise of the chosen resource
with a hypothesis is part of this mechanical work: when a hypothesis in the
current proof states exactly that premise, the compiler uses it
(\S\ref{subsec:liveness}).

Table~\ref{tab:resource-roles} places proof resources among the other objects
that appear in a proof step, using the running example of
\S\ref{sec:proof-state-compiler}.

\begin{table}[h]
  \centering
  \caption{Objects in a proof step and how \system{} treats them. Examples
  refer to the running example of \S\ref{sec:proof-state-compiler}.}
  \label{tab:resource-roles}
  \scriptsize
  \setlength{\tabcolsep}{3pt}
  \begin{tabular}{@{}>{\raggedright\arraybackslash}p{0.25\columnwidth}>{\raggedright\arraybackslash}p{0.24\columnwidth}>{\raggedright\arraybackslash}p{0.23\columnwidth}>{\raggedright\arraybackslash}p{0.2\columnwidth}@{}}
    \toprule
    \textbf{Object} & \textbf{Example} & \textbf{Role in a proof step} & \textbf{Handled by} \\
    \midrule
    Proof resource: lemma, axiom, local hypothesis
      & \texttt{Alossless\_F}
      & Used instead of new reasoning
      & Agent chooses; compiler finds and instantiates \\
    \addlinespace
    Hypothesis stating a premise of the chosen resource
      & \texttt{Ofll : islossless D2(O).O.f}
      & Fills that premise
      & Compiler \\
    \addlinespace
    Invariant, intermediate game, new lemma
      & A relational invariant for a loop
      & Must be invented
      & Agent \\
    \addlinespace
    Module, type
      & \texttt{D2(O).O} for the parameter \texttt{X}
      & Fills an argument of a resource
      & Compiler \\
    \addlinespace
    Program structure
      & The call to \texttt{guess} in the goal
      & Decides whether a resource applies now
      & Compiler tracks it \\
    \addlinespace
    Side condition
      & \texttt{islossless D2(O).O.f}, if no hypothesis states it
      & A goal that remains to be proved
      & Agent proves it \\
    \bottomrule
  \end{tabular}
\end{table}

\smallskip\noindent\textbf{Program structure decides which resources apply.}
Program structure is not a resource, but it determines which resources can be
used. In game-based proofs, many resources apply only at particular program
points: a lemma about a procedure call applies only while that call is still
visible in the goal, and a lemma about an oracle applies only at a call to that
oracle. Each accepted tactic changes the visible program structure, so a
resource may not apply yet, apply now, or no longer apply. For example, after
\texttt{inline*} replaces a call with the body of the procedure, a lemma about
that call no longer applies to the resulting goals; this is the unconscious
lowering of \S\ref{sec:overview}. For this reason, the compiler recomputes at
every proof state which resources apply, instead of treating the project's
lemmas as a fixed library.

\smallskip\noindent\textbf{What the agent sees.}
When the compiler shows a resource to the agent, it gives the name of the
resource, its statement instantiated at the current state, and which of its
arguments remain open (\S\ref{subsec:action}).
}

\begin{figure*}[t]
  \centering
  \resizebox{0.98\textwidth}{!}{\colorlet{archgray}{black!40}
\begin{tikzpicture}[
  x=1cm, y=1cm,
  font=\sffamily\footnotesize,
  band/.style={draw=black!14, fill=black!1.5, rounded corners=3pt, line width=0.5pt},
  bandlabel/.style={font=\sffamily\scriptsize\bfseries, text=black!62,
    anchor=west, align=left, text width=1.65cm, inner sep=0pt},
  box/.style={draw=#1, fill=#1!14, rounded corners=2pt, line width=0.6pt,
    anchor=south west, align=center, inner sep=2.5pt},
  flow/.style={-latex, line width=0.55pt, draw=black!55},
  lab/.style={font=\sffamily\scriptsize, text=black!60, inner sep=1.5pt},
]

\path[band] (0,7.5) rectangle (17.8,8.75);
\node[bandlabel] at (0.2,8.125) {Decomposer\\[-1pt]{\mdseries(\S\ref{sec:coordination})}};

\path[band] (0,4.55) rectangle (17.8,7.3);
\node[bandlabel] at (0.2,5.925) {Proof search\\[-1pt]{\mdseries(\S\ref{sec:tree-policy})}};

\path[band] (0,2.95) rectangle (17.8,4.35);
\node[bandlabel] at (0.2,3.65) {Proof-state compiler\\[-1pt]{\mdseries(\S\ref{sec:proof-state-compiler})}};

\path[band, draw=ablgrnC!70, fill=ablgrnA!22, line width=0.7pt] (0,0.3) rectangle (17.8,2.75);
\node[bandlabel, text=ablgrnC!55!black] at (0.2,1.525) {EasyCrypt interaction\\[-1pt]{\mdseries(trusted)}};

\node[box=ablgrnB, minimum width=4.8cm, minimum height=0.85cm] (jobs) at (2.1,7.7)
  {\textbf{Job scheduler}\\[-1pt]{\scriptsize runs lemma jobs in parallel;}\\[-2pt]{\scriptsize merges returned proofs}};
\node[box=ablgrnB, minimum width=8.9cm, minimum height=0.85cm] (dec) at (8.6,7.7)
  {\textbf{Decomposer} (LLM agent)\\[-1pt]{\scriptsize writes the intermediate games and lemma statements;}\\[-2pt]{\scriptsize assembles the proof of the main theorem}};
\draw[flow] (dec.west) -- node[lab, above] {lemma} (jobs.east);

\node[box=ablgrnB, minimum width=4.8cm, minimum height=2.1cm] (orch) at (2.1,4.85)
  {\textbf{Orchestrator}\\[2pt]
   {\scriptsize spawns agents at branch points}\\[-2pt]
   {\scriptsize prunes branches that fall behind}\\[-2pt]
   {\scriptsize records failed attempts}\\[-2pt]
   {\scriptsize selects the winning branch}};
\draw[flow] (3.9,7.7) -- node[lab, left] {lemma} (3.9,6.95);
\draw[flow] (5.1,6.95) -- node[lab, right] {proof or feedback} (5.1,7.7);

\draw[black!22, dashed, rounded corners=3pt, fill=white] (8.25,4.7) rectangle (17.6,7.05);
\draw[black!35, dashed, rounded corners=3pt, fill=white] (8.1,4.75) rectangle (17.45,7.1);
\node[font=\sffamily\scriptsize, text=black!50, anchor=north west, inner sep=0pt] at (8.25,6.98) {one per branch};
\node[box=ablbase, minimum width=3.7cm, minimum height=1.3cm] (mgr) at (8.3,4.95)
  {\textbf{Branch manager}\\[-1pt]{\scriptsize checks and dispatches each action;}\\[-2pt]{\scriptsize composes the next turn}};
\node[box=ablbase, minimum width=3.9cm, minimum height=1.3cm] (agent) at (13.4,4.95)
  {\textbf{LLM agent}\\[-1pt]{\scriptsize commit a tactic, undo, undo to a}\\[-2pt]{\scriptsize checkpoint, amend and replay,}\\[-2pt]{\scriptsize restart, or finish}};
\draw[flow] ([yshift=0.22cm]agent.west) -- node[lab, above] {action} ([yshift=0.22cm]mgr.east);
\draw[flow] ([yshift=-0.22cm]mgr.east) -- node[lab, below] {next turn} ([yshift=-0.22cm]agent.west);
\draw[flow] (orch.east |- 0,5.9) -- node[lab, above] {spawn,} node[lab, below] {prune} (8.1,5.9);

\node[box=ablours, minimum width=2.8cm, minimum height=0.8cm] (rba) at (9.2,3.2)
  {\textbf{Resource and}\\[-2pt]\textbf{binding analysis}};
\node[box=ablours, minimum width=1.7cm, minimum height=0.8cm] (ir) at (12.6,3.2)
  {\textbf{ProofIR}};
\node[box=ablours, minimum width=2.4cm, minimum height=0.8cm] (proj) at (14.9,3.2)
  {\textbf{State projection}};
\draw[flow] (proj.west) -- (ir.east);
\draw[flow] (ir.west) -- (rba.east);
\draw[flow] (rba.north) -- (rba.north |- mgr.south);
\node[lab, anchor=west] at (10.7,4.17) {applicable resources, realized tactics, diagnostics};

\node[box=ablgrnC, minimum width=4.8cm, minimum height=1.25cm] (final) at (2.1,0.6)
  {\textbf{Final check}\\[-1pt]{\scriptsize recompiles the proof from scratch}\\[-2pt]{\scriptsize in EasyCrypt; rejects any \texttt{admit}}};
\node[box=archgray, minimum width=4.0cm, minimum height=1.25cm] (ec) at (8.3,0.6)
  {\textbf{EasyCrypt session}\\[-1pt]{\scriptsize executes tactics; undo;}\\[-2pt]{\scriptsize checkpoint replay}};
\node[box=archgray, minimum width=4.0cm, minimum height=1.25cm] (log) at (13.3,0.6)
  {\textbf{Event log}\\[-1pt]{\scriptsize goals, errors, and accepted}\\[-2pt]{\scriptsize tactics, per proof state}};
\draw[flow] (8.7,4.95) -- node[lab, left] {tactic} (8.7,1.85);
\draw[flow] (ec.east) -- (log.west);
\draw[flow] (proj.south |- log.north) -- node[lab, right] {new state} (proj.south);
\draw[flow] (orch.south) -- node[lab, right] {winning proof} (final.north);
\end{tikzpicture}}
  \caption{\system{}'s architecture. The decomposer hands each lemma to the lemma prover as a job.
  Within a job, the orchestrator runs one LLM agent per branch of the proof tree:
  the agent chooses an action, the branch manager dispatches it to an EasyCrypt session,
  and the proof-state compiler turns the resulting state into the agent's next turn.
  Only the bottom band is trusted: a tactic changes the proof only if EasyCrypt accepts it,
  and a proof counts only after the final check recompiles it from scratch.}
  \label{fig:architecture}
\end{figure*}

\section{System Architecture}
\label{app:system-architecture}

Figure~\ref{fig:architecture} shows how \system{} is organized. The
architecture follows one rule: an LLM agent may choose proof actions, but it
does not own the proof state and does not decide when a proof is valid. The
figure has four bands, which also define the code categories reported in
Section~\ref{sec:evaluation}.

\smallskip
\noindent
\textbf{Decomposer.}
The decomposer is the LLM agent of Section~\ref{sec:coordination}. It writes
the intermediate games and lemma statements and submits individual lemmas as
jobs to a scheduler, which runs a bounded number of jobs in parallel. Each job
runs the lemma prover on one lemma and returns either a proof that has passed
the final check or feedback for the decomposer.

\smallskip
\noindent
\textbf{Proof search.}
Within a job, the orchestrator manages the proof tree of
Section~\ref{sec:tree-policy}. Each branch is extended by one LLM agent, paired
with a branch manager. The agent acts through six actions: commit a tactic,
undo the last step, undo to a checkpoint, amend an earlier tactic and replay,
restart, or finish. The branch manager checks each action, dispatches it to an
EasyCrypt session, and composes the agent's next turn from the current goal,
the outcome of the last action, and the output of the proof-state compiler.
The orchestrator observes all branches: it spawns agents at branch points,
prunes branches that fall behind, records failed attempts, and selects the
winning branch.

\smallskip
\noindent
\textbf{Proof-state compiler.}
After each action that changes the proof state, the compiler reads the new
state from the event log and runs the pipeline of
Section~\ref{sec:proof-state-compiler}: state projection, \proofir{}, and
resource and binding analysis. Its output, namely the applicable resources,
realized tactics, and diagnostics, becomes part of the agent's next turn. The
compiler only reads the proof state; it never changes it.

\smallskip
\noindent
\textbf{EasyCrypt interaction.}
The bottom band is the only part of the system that interacts with EasyCrypt,
and it is the trust boundary. An EasyCrypt session executes tactics, supports
undo and checkpoint replay, and records every goal, error, and accepted tactic
in an event log. A tactic changes the proof only if EasyCrypt accepts it. When a
branch finishes, the final check writes the proof into the source file,
recompiles the file from scratch in EasyCrypt, and rejects the proof if any
\texttt{admit} remains. Errors in the agents, the orchestrator, or the compiler
can therefore waste time, but they cannot cause an invalid proof to be
accepted.

\section{Dataset and Additional Case Studies}
\label{app:crypto-benchmark}

\subsection{Dataset Construction}
\label{app:dataset-construction}

Our benchmark collects proof tasks from EasyCrypt developments of
cryptographic schemes. It complements mathematical theorem-proving resources
such as LeanDojo~\cite{yang2023leandojo} with tasks involving probabilistic and
relational programs, adversaries and oracles, bad events, game hops, and
security reductions. We use a subset spanning different proof types and
difficulty levels for the ablation study in the main body. The public and
private parts of the benchmark are described below.

\smallskip\noindent\textbf{Building the dataset.}
Each task in the benchmark is a single lemma  
  taken from a real cryptographic development in EasyCrypt,
together with the surrounding context it needs (i.e., sibling definitions, cloned
theories, and imported lemmas). 
We remove the original proof; the task is to
reconstruct a proof script for the lemma, with its statement and context unchanged. Each lemma is labeled
with the scheme from which it comes, the cryptographic property it concerns, the proof
techniques it requires, and a difficulty level. 

\smallskip\noindent\textbf{The four parts and what each one measures.}
The benchmark is organized into four parts of increasing realism and difficulty.
They differ in where the proofs
come from, how hard they are, and crucially whether they could have been
seen during a model's pretraining.

\noindent\ding{182}~{\em Foundational primitives.} Textbook and building-block schemes such as
ElGamal and hashed-ElGamal, Schnorr, pseudorandom generators, private information
retrieval, Pedersen commitments, and the correctness of Cramer--Shoup. These are
short, local proofs, and they test whether the basic proof-state representation is
adequate: finite-set and list algebra, simple relational coupling, losslessness,
and single textbook reductions. This part establishes the floor of the benchmark
and isolates interface issues from genuine reasoning difficulty.

\noindent\ding{183}~{\em Deployed real-world schemes.} This part includes standardized
real-world schemes (e.g., ChaCha20-Poly1305, which we use as a case study).
Unlike Part \ding{182}, these are large, multi-lemma developments, so they test
whether an LLM agent can move beyond isolated lemmas and find useful proof
resources across lemmas, such as long game-hop chains, oracle
invariants, bad-event probability bounds, and reasoning that depends on context
established by other lemmas in the same project.

\noindent\ding{184}~{\em Post-quantum standards.} Lemmas from the machine-checked proofs of
the NIST post-quantum standards: the IND-CCA security of ML-KEM
(FIPS-203)~\cite{almeida2024mlkem,almeida2023kyber}, the tight security proof of
SLH-DSA/SPHINCS$^{+}$~\cite{barbosa2024sphincs}, the related XMSS
development~\cite{barbosa2023xmss}, and the security argument of ML-DSA/Dilithium.
These are the most demanding and most current proofs in the benchmark, addressing
the same problems that the standardization community itself verifies: lattice hardness
assumptions, the Fujisaki-Okamoto transform, and hash-based signature arguments.
Because these developments are recent and specialized, they also test whether an LLM agent can operate on proofs that lie well outside the well-trodden textbook
material.

\noindent\ding{185}~{\em A published case study with non-public proof scripts.} This part
consists of the CMAC/CBC-MAC family of proofs. We obtained the scripts directly
from their authors, who authorized their use in this evaluation. To our
knowledge, these scripts have not been publicly released. This reduces the
chance that a model has seen these particular proofs during training,
although it does not rule out exposure to related mathematics or proof
techniques.

\begin{table*}[t]
  \centering
  \caption{Final results for MEE-CBC and CMAC. For MEE-CBC, the decomposer
  proposed the helper lemmas for each final theorem; for CMAC, the lemma
  and equivalence statements come from the original development.}
  \label{tab:case-study-coverage}
  \small
  \setlength{\tabcolsep}{5pt}
  \renewcommand{\arraystretch}{1.2}
  \begin{tabular}{@{}>{\raggedright\arraybackslash}p{0.36\textwidth}
                      >{\raggedright\arraybackslash}p{0.32\textwidth}
                      >{\raggedright\arraybackslash}p{0.26\textwidth}@{}}
    \toprule
    Case study and scope & Completed proofs & Unfinished proofs \\
    \midrule
    MEE-CBC: confidentiality (\texttt{RCPA\_security}) &
    Final theorem and its 8 helper lemmas & None for this target \\
    \addlinespace
    MEE-CBC: plaintext integrity (\texttt{PTXT\_security}) &
    Final theorem and its 4 helper lemmas & None for this target \\
    \addlinespace
    MEE-CBC: functional correctness (\texttt{mee\_correct}) &
    Final theorem and its 21 helper lemmas & None for this target \\
    \addlinespace
    CMAC: 333 original statements, with syntax changes where needed &
    304: 299 with AI-generated proofs and 5 with existing proofs &
    29, including \texttt{CMAC\_security} \\
    \addlinespace
    CMAC: 11 reconstructed EMAC-to-CMAC statements &
    9, all with AI-generated proofs &
    2: \texttt{pr\_PRP\_PRF\_ECBC} and \texttt{CMAC} \\
    \bottomrule
  \end{tabular}
  \par\smallskip
  \begin{minipage}{\textwidth}
    \footnotesize
    MEE-CBC uses an existing verified CBC library. Of the 299 CMAC
    statements with AI-generated proofs, 239 have newly found proofs and
    60 reuse an AI-generated proof checked for another matching statement.
    CMAC lemma proofs are checked under the supplied dependencies, some of
    which remain unproved; the complete CMAC security proof is unfinished.
  \end{minipage}
\end{table*}

\smallskip\noindent\textbf{Sources and availability.}
Parts \ding{182}--\ding{184} are built from publicly available formal developments: the
\easycrypt{} example suite and standard libraries for the foundational and
deployed schemes, and the public machine-checked artifacts for the post-quantum
standards. 
For these parts, we will openly release our benchmark, except for the ML-DSA/Dilithium tasks, which we will release only with the permission of its authors. Part \ding{185} is different: the CMAC
development is private. Its proof scripts are not available in any public
repository, and we obtained them directly from the development's authors for the
purpose of this benchmark. We report aggregate CMAC results without releasing
the underlying private scripts. Readers seeking access should contact the
original authors.

\smallskip\noindent\textbf{Scope and difficulty.}
In total the benchmark comprises more than 1{,}600 lemma-level obligations across
these four parts, with the post-quantum developments contributing the largest
share. Within every part, obligations span three difficulty levels:
``easy'' obligations test local proof mechanics,
``medium'' ones require selecting and following a non-trivial proof strategy, and
``hard'' ones are the security-proof components---adversary and oracle reasoning,
game hops, and probability-bound composition. Because some of the developments
were written for older versions of the proof assistant, we port them to current
\easycrypt{} where feasible and record, per obligation, whether it is
machine-checkable today; the obligations that remain are flagged so that the
benchmark grows as more developments can be loaded and checked.

\subsection{Final Results for MEE-CBC and CMAC}
\label{app:case-study-coverage}

Table~\ref{tab:case-study-coverage} summarizes the completed proofs and the
remaining obligations in the two case studies. For MEE-CBC, we report the
three final theorem targets together with the helper lemmas that the
decomposer proposed for them. For CMAC, we report individual lemma and
equivalence statements, each counted once even if it was attempted more
than once. These counts refer to different proof tasks and are not compared
as success rates.

\smallskip\noindent\textbf{MEE-CBC.}
The agents receive the security
model and selected targets with their intermediate human lemmas removed.
Using Opus~5 for decomposition and Opus~4.8 for lemma proving, \system{} constructs
auxiliary lemmas and their proofs in separate runs for the selected tasks.
It completes all three final targets: confidentiality
(\texttt{RCPA\_security}), plaintext integrity (\texttt{PTXT\_security}), and
functional correctness (\texttt{mee\_correct}). Across the three targets,
the decomposer proposed 33 helper lemmas, and the lemma prover proved all 33
together with the three targets; each target run took 23--41 minutes.

The experiment reuses existing verified dependencies, including the CBC
library and its security proof. The agent-generated proofs are combined
with retained proofs in the final source files. Thus, the result establishes
the three target theorems without requiring the agents to regenerate every
proof in the supporting libraries. The final sources are available in the
\href{https://skyshannonprover.github.io/shannon-prover/results/\#case/mee}{MEE-CBC proof browser}.

\smallskip\noindent\textbf{CMAC.}
For CMAC, we evaluate lemma proving only (Phase~III): the lemma and
equivalence statements come from the original development, and we report
how many of them now have proofs that EasyCrypt accepts.
The original development contains 333 named statements, which we kept
unchanged except for updates to EasyCrypt syntax. Of these, 299 have
agent-generated proofs: 239 were proved directly, and 60 reuse a proof found
for a matching statement in another file. Five keep their existing proofs,
and 29 remain unfinished, including the final theorem \texttt{CMAC\_security}.
One file, which relates EMAC to CMAC, could not be loaded and referred to
missing definitions; we reconstructed it and report its 11 statements
separately: nine have agent-generated proofs and two remain unfinished.
All proofs are checked against the supplied dependencies, some of which
remain unproved. The counts therefore show how much of the proof work is
done, not a complete CMAC security proof; since lemmas differ in difficulty,
they also do not measure how much expert effort is saved.

\subsection{Expert Comments on MEE-CBC and CMAC}
\label{app:expert-comments}

We consulted
experts who participated in the developments or had substantial knowledge of them, and received their comments on
how LLM agents can accelerate their work.

\smallskip\noindent\textbf{MEE-CBC.} This development started from an existing
proof that CBC mode is secure with uniformly random
IVs, which was composed with a new proof that MAC-then-Encode-then-Encrypt is INT-PTXT secure. This new
proof was formalized by an expert EasyCrypt user with one
person-week of effort. A further proof refining the security
result to an implementation model (a C-like imperative
model in EasyCrypt) took an additional person-week from
the same expert. The main source of expert judgment in
this development is in choosing the right abstractions and
modularizing the proof; once this structure was fixed, much
of the remaining work was mechanical proof engineering.
Experts commented that the novel part of the MEE-CBC
proof involved little human insight, and would have been
automated with LLM agents.

\smallskip\noindent\textbf{CMAC.} The proof for CMAC took three person-months
by ordinary users of EasyCrypt. This effort also included
identifying and resolving small issues in the pen-and-paper
proof, as well as making nontrivial design decisions for
a part of the argument that could not be formalized as
written. (The formalized bound is looser than the pen-and-paper bound as a result.) The experts suggested that the
CMAC proof could have been greatly accelerated with LLM
agents, not only by reducing the amount of mechanical proof
work, but also by surfacing issues in the pen-and-paper proof
earlier and helping inform the high-level proof decisions that
affect the theorem statement.

\section{Baseline Failure Examples and Standout Agent Proofs}
\label{app:failures-and-standouts}

Baseline runs, in which the agent interacts directly with EasyCrypt, show
the three failure patterns of \S\ref{sec:overview}.

\noindent\ding{182}~\emph{Unconscious lowering: inlining before applying an equivalence.}
The agent recognized that the oracle equivalence \texttt{equ\_cc} should be
applied at a procedure call and recovered its quantified parameters. It then
used \texttt{inline *}. Although EasyCrypt accepted this tactic, it removed
the call structure needed to apply the equivalence. The agent spent
35 minutes on the resulting unsuccessful proof attempt. The baseline
interface did not warn that inlining would make the intended lemma
inapplicable.

\noindent\ding{183}~\emph{Semantic mislocalization: choosing a lemma by its name.}
The agent searched around \texttt{IFinRO}/\texttt{IndRO} and chose a lemma
whose name looked relevant. It spent more than 25 minutes trying to use
\texttt{pr\_RO\_FinRO\_D} and inlining eager and lazy loops. The useful next
step was to apply \texttt{byequiv} and reason about the calls to
\texttt{chacha\_enc2} and \texttt{poly\_mac2}. The agent followed name
similarity without matching the lemma to the procedure calls in the
current goal.

\noindent\ding{184}~\emph{Agent flinch: not applying an identified lemma.}
The agent repeatedly identified the right strategy: use \texttt{byequiv},
apply the bridge lemma \texttt{pr\_CCP\_OCCP}, and use the PRF distinguishing
argument. It proposed this plan four times but never applied the bridge
lemma. Instead, it tried local tactics such as \texttt{proc; inline *} and
\texttt{wp; skip; smt()}, or restarted the proof. The difficulty was in
supplying the arguments needed to apply the lemma. The baseline interface
did not provide the typed argument information needed to turn the plan into
a proof step.

\smallskip\noindent\textbf{Standout agent proofs.}
Agent proofs are sometimes better than the human proofs. For five ChaChaPoly
lemmas, the experts judged the agent proof better in concision, readability,
and semantic idiom (Fig.~\ref{fig:chachapoly-per-lemma}). We describe the two
marked with stars in the figure.

\emph{\texttt{chacha\_spec}: stating the loop invariant with the specification.}
The lemma states that the loop implementing ChaCha20 encryption computes the
functional specification \texttt{gen\_CTR\_encrypt\_bytes}, which encrypts the
message block by block with counters $1, 2, \ldots$. The difficulty is the loop
invariant. The human proof states it in terms of the number of blocks that
remain to be encrypted, so each iteration must show by integer arithmetic that
this number decreases by exactly one, with a case split on whether a full block
remains (23 lines). The agent instead states the
invariant with the specification itself: the ciphertext produced so far,
followed by the encryption of the remaining message from the current counter,
equals the encryption of the whole message. One iteration is then exactly the
existing lemma that removes one block from the encryption
(\texttt{gen\_CTR\_encrypt\_bytes\_cons}), and loop exit is the lemma for the
empty message (\texttt{gen\_CTR\_encrypt\_bytes0}). The arithmetic is still
needed, but it is already done inside these lemmas, so the proof takes 7 lines.

\emph{\texttt{step4\_badi}: describing the bad event exactly.}
This lemma is one step of a union bound in the integrity proof. Each call
to an internal procedure samples a fresh tag and compares it with a list of
earlier tags; the game records every compared pair in a list \texttt{lbad1},
and a collision in any pair is a bad event. To bound the probability of a
collision, the proof bounds each position $i$ of \texttt{lbad1} separately.
For this, it uses a modified game that, in the call that fills position $i$,
records in a flag \texttt{badi} whether the fresh tag equals the earlier tag
at that position. The lemma states that a collision at position $i$ in the
original game is no more likely than \texttt{badi} in the modified game, and
the proof relates the two games one call at a time.
The human proof (73 lines) carries an auxiliary invariant relating the
length of \texttt{lbad1} to the number of queries so far, and distinguishes
four cases for each call: no pair is added, or position $i$ was filled
earlier, is filled by this call, or is still empty afterwards.
The agent's proof (41 lines) states the relation between the two games
directly: before position $i$ is filled, \texttt{badi} is false, and after
it is filled, \texttt{badi} holds exactly when the pair at position $i$
collides. Each call then only needs to check whether it fills position $i$,
and the auxiliary invariant on the length of \texttt{lbad1} is no longer
needed.

\section{Ethics Considerations}
\label{app:ethics}

\noindent\textbf{Research scope.}
This work studies automated construction of formal cryptographic proofs. It
does not involve a recruited human-subject study, collect or process sensitive
personal data, or deploy experiments on third-party systems. It does not report
new vulnerabilities in deployed systems requiring coordinated disclosure.
The permissions for the cryptographic developments used in our evaluation,
including the non-public CMAC development, and the use of generative AI are
discussed in Section~\ref{sec:generative-ai-responsibility}.

\smallskip\noindent\textbf{Benefits and limitations of proof automation.}
The intended benefit of \system{} is to reduce the effort needed to obtain
machine-checked cryptographic proofs. A risk is that users may overinterpret
an accepted proof as an unconditional guarantee about a deployed system.
The guarantee is limited to the formal statement, its assumptions, and the
trusted verification infrastructure; it does not establish that a model
captures every property of an implementation. Users must therefore review
the specification and theorem statements, and check the final
proof scripts with EasyCrypt. An unsuccessful search also does not establish
that a theorem is false. These boundaries are essential to using proof
automation to support, rather than replace, judgments about cryptographic
security.

\end{document}